\documentclass[epsf,useAMS,usenatbib,usegraphicx]{mn2e}
\usepackage{epsf}
\usepackage{epsfig}
\usepackage{aas_macros}
\usepackage{amsmath}
\usepackage{url}
\usepackage{appendix}
\usepackage{deluxetable}
\usepackage[usenames]{color}
\usepackage{multirow}
\usepackage[table]{xcolor}
\definecolor{lightgray}{gray}{0.9}

\newcommand{\ion}[2]{#1\,{\small #2}}
\newcommand{\kepler} {\emph{Kepler}}

\newcommand{\amlt} {\alpha_{\rm MLT}}
\newcommand{\dov} {d_{\rm ov}}
\newcommand{\eqn} [1] {
\begin{equation}
#1
\end{equation}}

\title[KIC\,11754974]{Asteroseismology of KIC\,11754974: a high-amplitude SX\,Phe pulsator in a 343-day binary system}

\author[Simon J. Murphy et al.]
{S. J. Murphy$^{1,2}$\thanks{email: smurphy6@uclan.ac.uk}, A. Pigulski$^{3}$, D.W. Kurtz$^{1}$, J.C. Su\'arez$^{4}$, G. Handler$^{5}$, \and L.A. Balona$^{6}$, B. Smalley$^{7}$, K. Uytterhoeven$^{8,9}$, R. Szab\'o$^{10}$, A.O. Thygesen$^{11,12}$,  \and V. Elkin$^{1}$, M. Breger$^{13,14}$, A. Grigahc\`ene$^{4}$, J.A. Guzik$^{15}$, J.M. Nemec$^{16}$ and \and J. Southworth$^{7}$.\\
\\
$^{1}$Jeremiah Horrocks Institute, University of Central Lancashire, Preston PR1 2HE\\
$^{2}$Centro de Astrof\'isica, Faculdade de Ci\^encias, Universidade do Porto, Rua das Estrelas, 4150-762 Porto, Portugal\\
$^{3}$Instytut Astronomiczny, Uniwersytet Wroc\l{}awski, Kopernika 11, 51-622 Wroc\l{}aw, Poland\\
$^{4}$Instituto de Astrof\'isica de Andaluc\'ia, Camino Bajo de Hu\'etor 50, Granada, 18008, Spain\\
$^{5}$Nicolaus Copernicus Astronomical Center, Bartycka 18, 00-716 Warsaw, Poland\\
$^{6}$South African Astronomical Observatory, PO Box 9, Observatory 7935, Cape Town, South Africa\\
$^{7}$Astrophysics Group, Keele University, Staffordshire, ST5 5BG\\
$^{8}$Instituto de Astrof\'isica de Canarias, 38200 La Laguna, Tenerife, Spain\\
$^{9}$Dept. Astrof\'{\i}sica, Universidad de La Laguna, 38206 La Laguna, Tenerife, Spain\\
$^{10}$Konkoly Obs., Research Centre for Astronomy and Earth Sciences, HAS, H-1121 Budapest, Konkoly Thege Mikl\'{o}s \'{u}t 15-17, Hungary\\
$^{11}$Zentrum F\"{u}r Astronomie der Universit\"{a}t Heidelberg, Landessternwarte, K\"{o}nigstuhl 12, 69117 Heidelberg, Germany\\
$^{12}$Nordic Optical Telescope, Apartado 474, E-38700 Santa Cruz de La Palma, Santa Cruz de Tenerife, Spain\\
$^{13}$Institut f\"ur Astronomie, T\"urkenschanzstra\ss{}e 17, 1180 Wien, Austria\\
$^{14}$Department of Astronomy, University of Texas, Austin, TX 78712, USA\\
$^{15}$Los Alamos National Laboratory, XTD-2 MS T-086, Los Alamos, NM 87545-2345, USA\\
$^{16}$Department of Physics and Astronomy, Camosun College, Victoria, British Columbia V8P 5J2, Canada\\}

\begin{document}

\maketitle

\begin{abstract}
The candidate SX\,Phe star KIC\,11754974 shows a remarkably high number of combination frequencies in the Fourier amplitude spectrum: 123 of the 166 frequencies in our multi-frequency fit are linear combinations of independent modes. Predictable patterns in frequency spacings are seen in the Fourier transform of the light curve. We present an analysis of 180\,d of short-cadence \textit{Kepler} photometry and of new spectroscopic data for this evolved, late A-type star. We infer from the1150-d, long-cadence light curve, and in two different ways, that our target is the primary of a 343-d, non-eclipsing binary system. According to both methods, the mass function is similar, $f(M)=0.0207 \pm 0.0003$\,M$_{\sun}$. The observed pulsations are modelled extensively, using separate, state-of-the-art, time-dependent convection (TDC) and rotating models. The models match the observed temperature and low metallicity, finding a mass of 1.50--1.56\,M$_{\sun}$. The models suggest the whole star is metal-poor, and that the low metallicity is not just a surface abundance peculiarity. This is the best frequency analysis of an SX\,Phe star, and the only \textit{Kepler} $\delta$\,Sct star to be modelled with both TDC and rotating models.
\end{abstract}


\section{Introduction}

The goal of the \textit{Kepler} mission is to find Earth-like planets in the habitable zone \citep{kochetal2010}. The methodology inherent in this task is the simultaneous, space-based observation of $\sim$150\,000 stars in a 115\,deg$^2$ field of view for transiting events.

\textit{Kepler} observations are made in two cadences: most observations take place in long cadence (LC) format, where photometric readouts are co-added for 29.4\,min. Short cadence (SC) observations have effective integration times of 58.8\,s. These cadences make the data excellent for asteroseismic investigations. \textit{Kepler} data are organised into quarters, denoted Q$n$, and SC data are further subdivided into three one-month segments, denoted Q$n.m$.

The success of the \textit{Kepler} mission lies in its unprecedented $\mu$mag photometric precision. Even for a star as faint as KIC\,11754974 (Kp = 12.7\,mag), the subject of this work, the average noise amplitude in the Fourier transform is $\sim$5\,$\mu$mag at low frequency, decreasing gradually to $\sim2$\,$\mu$mag for frequencies above about 70\,d$^{-1}$.

\textit{Kepler} photometry not only offers higher precision, but also provides nearly-continuous observations. Progress in the analysis of $\delta$\,Sct stars is no longer hampered by the insurmountable time gaps in the data associated with ground-based observations. Most stars now have almost continuous observations in LC mode from the commissioning run of Q0 right through to the most recent quarter at the time of writing (Q13), spanning over 1150\,d; the same is true for some targets in SC mode. This long time-span is particularly important for resolving peaks that are very closely spaced in frequency, and for studies of frequency or amplitude modulation, among other reasons (see, e.g., \citealt{murphy2012b}). Completely alleviating Nyquist ambiguities with \kepler\ data can now also be added to that list \citep{murphyetal2012b}.

The $\delta$\,Sct stars are located at the junction of the classical Cepheid instability strip and the main-sequence, with pulsation periods between 18\,min and 8\,hr. They are driven by the opacity mechanism in the \ion{He}{II} partial ionisation zone. SX\,Phe stars are the Population\,II counterparts of the $\delta$\,Sct stars, characterized by low metallicity and high tangential velocities \citep{nemec&mateo1990, balona&nemec2012}. Within the last decade, observations of SX\,Phe stars have revealed that they are not necessarily intrinsically high-amplitude pulsators \citep{olechetal2005, balona&nemec2012}, but show amplitude distributions similar to the Population\,I $\delta$\,Sct stars, viz., having peak pulsation amplitudes of a few mmag (see \citealt{uytterhoevenetal2011} for a statistical study). Traditionally, $\delta$\,Sct stars with amplitudes above 0.3\,mag were termed HADS (High Amplitude Delta Scuti) stars, and were found to pulsate mostly in one or two radial modes. Their high amplitudes are attributed to their evolved stages (e.g. \citealt{petersen&dalsgaard1996}, who describe HADS stars as being in the immediate post-MS stage). KIC\,11754974 has a maximum peak-to-peak light variation of only 0.24\,mag; based on tangential velocities was identified as an SX\,Phe star by \citet{balona&nemec2012} and was the highest-amplitude \kepler\ SX\,Phe star they identified. In this work, we find KIC\,11754974 to have properties in common with both HADS and SX\,Phe stars.

\section{Binarity of KIC\,11754974}
\label{sec:binarity}
\subsection{Doppler-shifted frequencies}

\begin{figure}
\centering
\includegraphics[width=85mm]{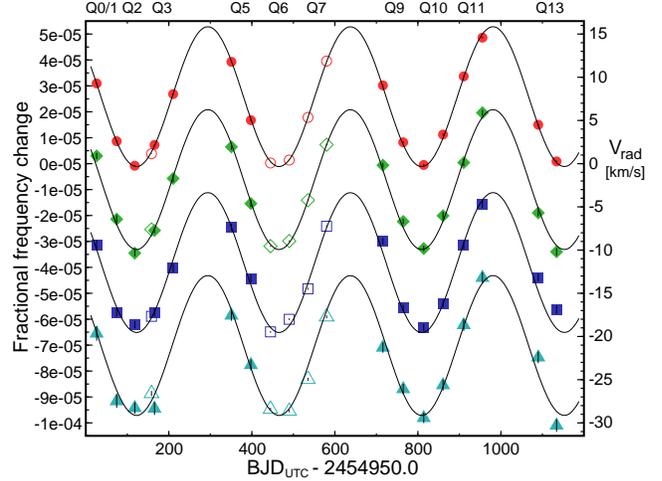}
 \caption{Temporal changes in the frequency of the four strongest modes excited in KIC\,11754974, $f_1$, $f_2$, $f_3$ and $f_4$, top to bottom. The values derived from LC and SC data are plotted with filled and open symbols, respectively. The sinusoid with a period of 343\,d fitted to the $f_1$ data is also shown. For clarity, an offset has been added to the $f_2$, $f_3$ and $f_4$ data; the sinusoid has also been shifted accordingly. The right-hand side ordinate shows radial velocities calculated from the Doppler effect.}
\label{dopp}
\end{figure}

Our target falls on \textit{Kepler}'s failed Module\:3, meaning it is only observed for three-quarters of each year. For our frequency analysis we utilized the higher sampling of the SC data, the longest uninterrupted run of which is Q6 and Q7, covering about 180\,d. In the course of analysing the combined Q6+Q7 dataset it appeared that in Fourier spectra a significant residual signal occurs near the strongest terms after their subtraction. This indicated frequency and/or amplitude changes which prompted us to carry out the analysis separately for shorter datasets. Both LC and SC data were used in this analysis; they were split into 30--50\,d subsets. For each subset, the frequency set containing the 72 strongest terms (20 independent and 52 harmonics and combinations) was fitted by means of non-linear least-squares. The resulting frequencies of the strongest modes in different subsets were then compared. The result is shown in Fig.\,\ref{dopp}, where we plot the fractional frequency change of the four frequencies with the highest amplitudes in a given time subset, relative to their amplitudes in the arbitrarily chosen reference subset, here the SC Q6 data. As can be seen, the frequencies of all four of the strongest modes behave in the same manner and can be described by a sinusoid with a period of about 343\,d.

The only plausible explanation of the changes seen in Fig.\,\ref{dopp} is the motion of KIC\,11754974 in a binary system. The frequency changes can be interpreted in terms of the Doppler effect and transformed to radial velocities via the equation $$V_{\rm rad} = c \frac{f_i - f_{\rm ref}}{f_{\rm ref}},$$ where $c$ represents the speed of light. These radial velocities can be used to derive parameters of the spectroscopic orbit of the pulsating primary. It can be seen from Fig.\,\ref{dopp} that the peak-to-peak radial velocity of KIC\,11754974 amounts to about 16\,km\,s$^{-1}$.

In order to derive parameters of the spectroscopic orbit, however, we proceeded in another way. Instead of using radial velocities calculated from frequencies, we decided to use the $O-C$ diagram for the times of maximum light. The advantage of using the $O-C$ diagram is that in this diagram the changes of period accumulate. For large orbital-to-pulsation period ratios, spectroscopic elements can be derived from the $O-C$ diagram more accurately than when using radial velocities calculated from the Doppler effect.

\begin{figure*}
\centering
\includegraphics[width=135mm]{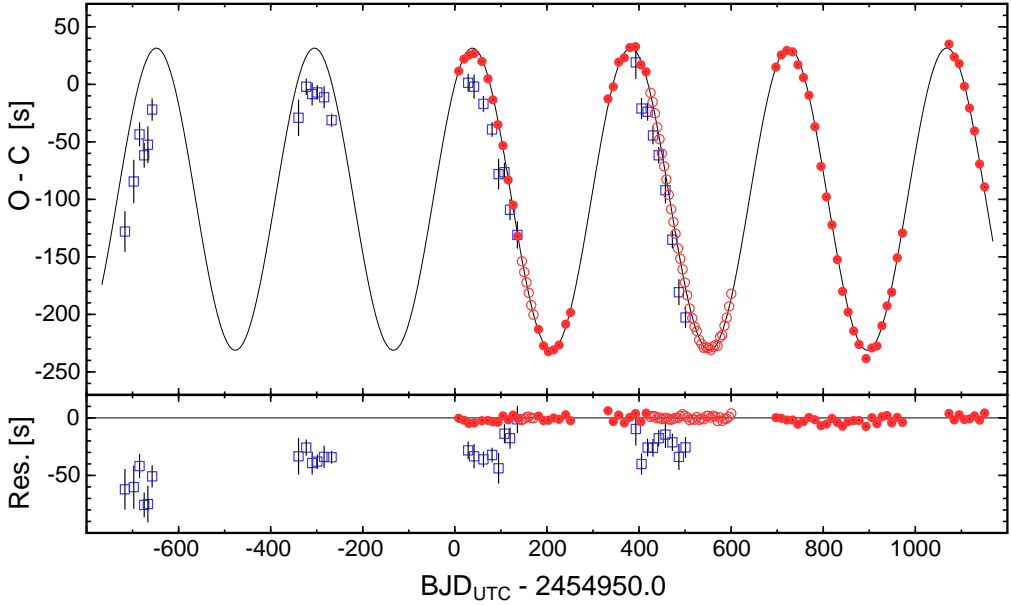}
 \caption{The $O-C$ diagram for the times of maximum light of the strongest mode of KIC\,11754974 (upper panel). Filled and open circles denote \textit{Kepler} LC and SC data, respectively. The open squares are the 2007--2010 WASP data. The lower panel shows the residuals from the eccentric orbit fit, shown as a continuous line in the upper panel. The parameters of the fit are given in Table\,\ref{eccfit}.}
\label{ocd}
\end{figure*}

Since the amplitude of the strongest mode is several times larger than the amplitudes of the next three modes, we decided to use only the times of maximum light of the strongest mode. First, all terms with amplitudes higher than 40\,ppm except for $f_1$ and its harmonics were removed from the data. The residuals were divided into 5--20\,d non-overlapping subsets. For each subset, a sinusoid with frequency $f_1$ and its harmonics were fitted in order to obtain amplitude and phase information. The time of maximum light was calculated from the phase of $f_1$ with the epoch adopted to be the nearest to the average time of observation in a given subset. The resulting observed ($O$) times of maximum light were compared to the calculated ($C$) times of maximum light according to the following ephemeris:
$$C(E) = \mbox{2454950.021179} + \mbox{0{\fd}0611817444} \times E,$$
where $E$ is the number of cycles elapsed from the initial epoch, in BJD using Coordinated Universal Time (UTC; \citealt{eastmanetal2010}). As such, our \kepler\ times do not suffer the timing error discovered in 2012 November\footnote{\url{http://archive.stsci.edu/kepler/timing_error.html}}.

The Wide Angle Search for Planets (WASP) is a multi-site, multi-camera imaging system obtaining photometric data with an accuracy better than 1\:per\:cent for objects with $7 \leq V \leq 11.5$ \citep{pollaccoetal2006}. Lower-quality data are obtained for stars with $11.5 \leq V \leq 15$. Although the \textit{Kepler} data have much higher precision and duty cycle, the WASP observations have time-spans extending back to years before \textit{Kepler} came online. We utilised this longer time-base of the WASP data to confirm our $O-C$ results, the values of which are plotted in Fig.\,\ref{ocd}.

Using the more precise \textit{Kepler} times of maximum light (for both LC and SC data), we fitted the $O-C$ parameters assuming an eccentric orbit. The parameters of the fit are given in Table\,\ref{eccfit}. The meaning of the parameters is standard: $T_0$ denotes the epoch of periastron passage; $e$, eccentricity; $\omega$, longitude of periastron; $K_1$, half-range of the primary's radial velocity variation; $P_{\rm orb}$, the orbital period; $P_0$, the pulsation period; $a_1$, the semi-major axis of primary's absolute orbit; and $f(M)$, the mass function. The residuals from the fit are equal to 3.28\,s for the LC and 1.47\,s for the SC data. At least part of the residual scatter comes from the small-amplitude variation due to modes that were not subtracted from the data.

The lower panel of Fig.\,\ref{ocd} indicates the WASP data have large residuals (in time) when compared with the \textit{Kepler} data. We tested whether the shift is intrinsic by comparing two stars in which such shifts were deemed to be unlikely -- we chose two relatively bright, short-period W\,UMa stars, namely: KIC\,8554005 and KIC\,9392683 (1SWASPJ191805.45+444115.4 and 1SWASPJ190413.72+455657.6, respectively). We investigated the strongest (second) harmonic in the Fourier transform of each star, and define a parameter DT = $T_{\rm max}$(WASP) - $T_{\rm max}$(Kepler), where $T_{\rm max}$ represents the time of maximum light closest to the mean epoch of a given dataset. For KIC\,8554005 we determine DT $= -5.8\,\pm\,5.2$\,s and for KIC\,9392683 DT $= -13.7\,\pm\,6.0$\,s, and note that DT is therefore zero to within 1--2\,$\sigma$. For KIC\,11754974 we found DT $= -34.2\,\pm\,2.8$\,s from all data or $-27.2\,\pm\,2.1$\,s if the first season is excluded. We therefore conclude that the shift for KIC\,11754974 is intrinsic and probably occurs because the WASP and \textit{Kepler} bands are different; the central wavelengths are $\sim$550\,nm for WASP and $\sim$660\,nm for \textit{Kepler}. We note that WASP and \textit{Kepler} use heliocentric and barycentric Julian date, respectively, and that contributes up to a couple of seconds to the offset.

\begin{table}
\centering
\caption{Parameters of the spectroscopic orbit derived from the $O-C$ diagram. $T_0$ is given in BJD$_{\rm UTC}$.}
\label{eccfit}
\begin{tabular}{cc}
\hline
Parameter & Value \\
\hline
$T_0$ & \phantom{1}2454999 $\pm$ 37\phantom{000000}\\
$e$ & \phantom{1}0.013\phantom{$\pm$}$^{+0.011}_{-0.006}$\phantom{1}\,\\
$\omega$ [$^\circ$] & 102 $\pm$ 38\phantom{1}\\
$K_1$ [km s$^{-1}$] & 8.35 $\pm$ 0.04\\
$P_{\rm orb}$ [d] & 343.27 $\pm$ 0.34\phantom{00} \\
$P_0$ [d] & 0.0611817444 $\pm$ 0.0000000011\\
$a_1\sin{i}$ [AU] & 0.2634 $\pm$ 0.0013\\
$f(M)$ [M$_\odot$] & 0.02069 $\pm$ 0.00031\\
\hline
\hline
\end{tabular}
\end{table}

One can see that the orbit has a small eccentricity. The secondary is likely to be a low-mass main-sequence star. Assuming the primary's mass lies in the range 1.5--2.5\,M$_\odot$, we get $M_2=$~0.42--0.58\,M$_\odot$ for inclination $i=$~90$^\circ$ and $M_2=$~0.65--0.87\,M$_\odot$ for $i=$~45$^\circ$. This indicates a K or early M-type secondary companion. Its contribution to the total flux is therefore not larger than a few per cent unless the inclination is very low. This reduces the possibility of detection of variability of the secondary if present at all.

Having calculated the orbital elements we have corrected the BJD times for short-cadence Q6 and Q7 data for the light-time effect in the binary system. Because in different quarters different detectors and apertures are used, it is expected that fluxes from different quarters are not directly comparable and need to be scaled. Using the amplitudes of the strongest modes we found that the amplitudes derived from Q7 data are larger by a factor of 1.0087 than those obtained from Q6 data. Therefore, the Q7 data were divided by this factor prior to combining. Subsequently, this high-duty cycle combined Q6+Q7 SC dataset was subject of time-series analysis (Section\,\ref{photometry}), where we have used only these adjacent SC quarters to avoid aliasing difficulties.

\subsection{Frequency modulated stars}

The effect of the binary system on light arrival times is that of frequency modulation for the pulsation modes. \citet{shibahashi&kurtz2012} showed how  a frequency multiplet is induced around each pulsation mode as a result of the binarity. The frequency spacing can be used to infer $P_{\rm orb}$, and the amplitudes and phases of the frequency multiplet give $T_0$, $e$ and $f(M)$, provided the dataset is longer than $P_{\rm orb}$. $f(M)$ is calculable using the ratio of the amplitudes of the first orbital sidelobes to those of the central pulsation frequency \citep{shibahashi&kurtz2012}. We used this method to check our binary parameters, and we will show that the results we obtain using this method are entirely consistent with those derived from the $O-C$ diagram. We applied this methodology to the \textit{Kepler} LC Q0--13 dataset, spanning 1153\,d (3.36 orbits).

We started with the highest-amplitude peak in the Q0--13 dataset, which lies at 16.344745\,d$^{-1}$ with an amplitude of 51.79\,mmag. We extracted this peak and its orbital sidelobes and applied non-linear least-squares fitting routines to improve the frequencies, amplitudes and phases. From the mean spacing of the multiplet we derive $P_{\rm orb} = 343.66\,\pm\,0.48$\,d, which is consistent with that derived via the $O-C$ diagram. Given the slightly lower uncertainty on $P_{\rm orb}$ from the latter method, we used the ($O-C$)-calculated value of $P_{\rm orb}$ to force-fit the sidelobes to be exactly equally spaced. We chose the zero point in time, such that the sidelobes have the same phase with respect to each other, but are $\pi/2$ out of phase with the central frequency, thus demonstrating the triplet is orbital in nature. Table\,\ref{table:FM} displays the results for this frequency and three others, each of which shows a frequency triplet split by the orbital frequency.

\begin{table*}
\centering
\caption[]{A least-squares fit of the frequency triplets for the four highest-amplitude modes. The sidelobes are fixed at frequencies of $\nu_{\rm osc} \pm \nu_{\rm orb}$. The zero point in time, $t_0$ = BJD$_{\rm UTC}$\,24555331.39590, was chosen so that the sidelobes have equal phases and the difference in phase between the sidelobes and central peak is $\pi/2$. Column 4 thus shows that the difference in phases of the sidelobes with each other is zero within the errors, and Column 5 demonstrates their phase difference from the central peak is $\pi/2$ within the errors. Column 6 contains the amplitude ratios of the sidelobes to the central frequency. Since these values are small compared to unity, they are approximately equal to $\alpha$, the amplitude of the phase modulation. Column 7 shows the expected theoretical outcome that the ratio of $\alpha$ to the oscillation frequency is the same for all triplets. We note that for the first triplet, the amplitudes of the sidelobes are only equal at the $4\sigma$ level, and that the errors provided are overestimated because there are still many frequencies left in the data.}
\small
\begin{tabular}{ccccccc}
\hline
\multicolumn{1}{c}{frequency} & \multicolumn{1}{c}{amplitude} &   
\multicolumn{1}{c}{phase} & \multicolumn{1}{c}{$\phi_{+1}-\phi_{-1}$} & 
\multicolumn{1}{c}{${\langle\phi_{+1}-\phi_{-1}\rangle}-{\phi_0}$} & 
\multicolumn{1}{c}{${\alpha=(A_{+1}+A_{-1})}/{A_0}$} & 
$\alpha/\nu_{\rm osc}$ 
\\
\multicolumn{1}{c}{d$^{-1}$} & \multicolumn{1}{c}{mmag} &   
\multicolumn{1}{c}{radians} & \multicolumn{1}{c}{radians} & 
\multicolumn{1}{c}{radians} &  & 
$\times 10^{-3}$\,d 
\\
\hline
16.341823	&$\phantom{1}	3.964	\pm	0.048	$&$	-1.6554	\pm	0.0120	$&				&				&				&				\\
16.344745	&$	51.792	\pm	0.049	$&$	-0.0545	\pm	0.0009	$&$	-0.011	\pm	0.017	$&$	-1.606	\pm	0.008	$&$	0.157	\pm	0.003	$&$	9.58	\pm	0.16	$\\
\vspace{2mm}16.347667	&$\phantom{1}	4.148	\pm	0.048	$&$	-1.6663	\pm	0.0115	$&				&				&				&				\\
21.396067	&$\phantom{1}	1.116	\pm	0.048	$&$\phantom{-}	0.9328	\pm	0.0427	$&				&				&				&				\\
21.398989	&$	11.046	\pm	0.049	$&$\phantom{-}	2.5642	\pm	0.0044	$&$	\phantom{-}0.012	\pm	0.060	$&$	-1.626	\pm	0.030	$&$	0.204	\pm	0.012	$&$	9.54	\pm	0.58	$\\
\vspace{2mm}21.401911	&$\phantom{1}	1.140	\pm	0.048	$&$\phantom{-}	0.9443	\pm	0.0418	$&				&				&				&				\\
20.904493	&$\phantom{1}	0.766	\pm	0.048	$&$\phantom{-}	0.0139	\pm	0.0623	$&				&				&				&				\\
20.907414	&$\phantom{1}	8.320	\pm	0.049	$&$\phantom{-}	1.6598	\pm	0.0059	$&$	\phantom{-}0.045	\pm	0.086	$&$	-1.623	\pm	0.043	$&$	0.189	\pm	0.016	$&$	9.02	\pm	0.78	$\\
\vspace{2mm}20.910336	&$\phantom{1}	0.803	\pm	0.048	$&$\phantom{-}	0.0589	\pm	0.0595	$&				&				&				&				\\
20.940632	&$\phantom{1}	0.584	\pm	0.048	$&$\phantom{-}	2.1123	\pm	0.0818	$&				&				&				&				\\
20.943554	&$\phantom{1}	5.890	\pm	0.049	$&$	-2.5581	\pm	0.0083	$&$	-0.003	\pm	0.114	$&$	-1.614	\pm	0.058	$&$	0.202	\pm	0.023	$&$	9.62	\pm	1.10	$\\
20.946475	&$\phantom{1}	0.603	\pm	0.048	$&$\phantom{-}	2.1090	\pm	0.0792	$&				&				&				&				\\
\hline
\hline
\end{tabular}
\label{table:FM}
\end{table*}

The mass function is calculable as: $$f(m_1, m_2, \sin i) = \left(\frac{A_{+1} + A_{-1}}{A_0}\right)^3 \frac{P_{\rm osc}^3}{P_{\rm orb}^2} \frac{c^3}{2 \pi G}$$ where the amplitudes of the central, higher and lower frequency components of the triplet are denoted $A_0$, $A_{+1}$ and $A_{-1}$, respectively; $P_{\rm osc}$ is the oscillation period; and all quantities are in SI units. Via this method we find $f(M) = 0.02069 \pm 0.00036$\,M$_{\sun}$, which is exactly the value calculated via the $O-C$ method. Eq.\,23 of \citet{shibahashi&kurtz2012} allows one to calculate $a_1 \sin i$, too, as $$a_1 \sin i = \frac{P_{\rm osc}}{2\pi} \alpha c.$$ Application of this formula yields $a_1 \sin i = 0.2629 \pm 0.0051$\,AU, in good agreement with the $O-C$ value.

That such information is discernible from the Fourier transform highlights the importance of having continuous long-term observations available for $\delta$\,Sct stars, whose high-amplitude, high-frequency pulsations produce sidelobes with high signal-to-noise (S/N).

We also looked for eclipses in the SC data at the times when the stars are aligned to the line of sight. The possibility of finding such a shallow eclipse in the residuals of the light curve of a binary with this long a period, even when our multi-frequency fit was subtracted, was very small. Although we did not find an eclipse, for such a wide system this provides only a slight constraint on the orbital inclination (an upper limit that is near 90$^{\circ}$).

\section{Spectroscopic observations}
\label{sec:spectroscopy}
We obtained three spectra of KIC\,11754974 with the FIES spectrograph at the 2.5-m Nordic Optical telescope (NOT). The star is too faint for high resolution observations with these facilities so to obtain spectra of a higher S/N we observed with medium resolution (R=25000). Even still, observing conditions rendered the S/N ratio of the obtained spectra low -- the S/N ratio of the averaged spectrum, estimated from the noise level of reduced 1D spectra, is about 45 in the spectral region of H$\alpha$ and about 40 near H$\beta$. The difficulties in determining the right continuum level limit the accuracy of our determinations, which are provided to offer some constraints on fundamental parameters, rather than as definite determinations. We stress that higher resolution, higher S/N spectra are still needed.
 
The Balmer-line profiles are good indicators of effective temperature for A-type stars, thus we have compared observed and synthetic profiles of the H$\alpha$ and H$\beta$ lines. The synthetic calculations of Balmer profiles were done using the {\small SYNTH} code by \citet{piskunov1992} and model atmospheres from \citet{heiteretal2002b} with solar metallicity (as determined by \citealt{anders&grevesse1989}). The source of the spectral line list was the Vienna Atomic Line Database (VALD; \citealt{kupkaetal1999}). Later analysis of metal lines demonstrated metal deficiency with respect to solar abundances, so we switched to working with low metallicity models ([M/H] = $-0.5$). For H$\alpha$ and H$\beta$ the best fits for observed and synthetic profiles were obtained for this metallicity model, yielding $T_{\mathrm{eff}}$ = 7000\,K and 6800\,K, respectively. Considering possible errors in continuum determination of several percent, we estimate from the Balmer lines that $T_{\mathrm{eff}}$ = $7000 \pm 200$\,K.

With our spectral material we cannot make reliable determinations of other parameters, namely $\log g$ and the microturbulence velocity $v_{\rm mic}$. We calculated a number of synthetic spectra keeping $T_{\mathrm{eff}}$ and one other parameter fixed, (thus permitting the third variable to change) to find the best fit of the metal lines. After several iterations we accepted a model atmosphere with $T_{\mathrm{eff}}$ = 7000\,K, $\log g$ = 3.6 (cgs) and did calculations for microturbulence velocity resulting in $v_{\rm mic} =2$\,km\,s$^{-1}$. This $\log g$ value is only loosely constrained and we considered that the star shows HADS-star characteristics and is probably more evolved. However, the $\log g$ value obtained is lower than we expected, given that cool $\delta$\,Sct stars tend to pulsate in low overtones \citep{breger&bregman1975}, and a low overtone mode at 16\,d$^{-1}$ (the frequency of our highest-amplitude mode) implies an evolutionary status much closer to the ZAMS. We compared the spectrum with the spectrum of $\delta$ Sct itself, taken from the Elodie archive (\url{http://atlas.obs-hp.fr/elodie/}). The shapes of the H$\beta$ profiles are very similar in these stars. However, the spectrum of $\delta$ Sct is richer in metal lines.
 
The weak spectral lines in KIC\,11754974 are lost in the noise. In the averaged spectrum we identified lines of \ion{Fe}{I}, \ion{Fe}{II}, \ion{Ti}{II}, \ion{Mg}{I}, \ion{Mg}{II} and some others. For identification we calculated a synthetic spectrum using our best fitting parameters. From several lines including \ion{Fe}{II} $\lambda$5018.440, \ion{Mg}{I} $\lambda$5172.684, \ion{Mg}{I} $\lambda$5183.604 we determined a projected rotational velocity: $v \sin i = 31 \pm 2$\,km\,s$^{-1}$. A more cautious interpretation, taking into account more lines (with a poorer overall fit) is $v \sin i = 25 \pm 7$\,km\,s$^{-1}$ -- given the low S/N of our spectrum, we adopt the more cautious $v\sin i$ value.

Using several lines we determined the abundances of iron: $\log\,N/N_{\rm tot}  = -5.8 \pm 0.3$ (while solar is $-4.54$)\footnote{as indicated in logarithmic form in the VALD.}, and for magnesium: $\log\,N/N_{\rm tot} = -5.9 \pm 0.3$ (solar is $-4.46$). We note that fitting H$\alpha$ and H$\beta$ with [M/H] = $-1.0$ yielded negligible differences from the [M/H] = $-0.5$ models. Many other chemical elements in the spectra also show significant deficiencies but the lines are blended or below the detection threshold and cannot be tested. There is a problem in that many lines of \ion{Fe}{I} that are not visible in the observed spectra appear in synthetic ones and require yet lower abundances for best fitting than we obtained from available lines, the latter mostly belonging to \ion{Fe}{II}. One must decrease the effective temperature to reduce such a disequilibrium. At least one strong available line of \ion{Fe}{I} 4404.750\,\AA\ shows similar abundances to the \ion{Fe}{II} lines. This problem needs higher quality spectra for further analysis. Str\"omgren photometry is desirable for independent determinations of $T_{\mathrm{eff}}$ and log\,$g$ parameters. Our parameter determination from spectroscopy is given in Table\,\ref{spectroscopic-parameters}.

\begin{table}
\setlength{\extrarowheight}{3pt}
\centering
\caption{Atmospheric parameters determined from spectroscopy.}
\begin{tabular}{c c c c c}
\hline
\vspace{-1mm}T$_{\mathrm{eff}}$ & $\log g$ & [Fe/H] & $v\sin i$ & $v_{\rm mic}$\\ 
K & (cgs) & dex & km\,s$^{-1}$ & km\,s$^{-1}$ \\
\hline\vspace{1mm}
$7000\pm200$ & $3.6 \pm 0.3$ & $-0.5^{+0.2}_{-0.5} $& $25 \pm 7$ & 2\\
\hline
\hline
\end{tabular}
\label{spectroscopic-parameters}
\end{table}

With these restrictions in mind, verification of the low metallicity was provided by a semi-automated routine for 476 lines from the VALD database in the 5000--5200\,\AA\ wavelength range. The best fit [M/H] value using $\chi^2$ minimisation was $-0.58$\,dex, when the $\log g$ value was held at the determined value of 3.6. The [M/H] value is not the [Fe/H] value, but is close to it.

An independent check on our spectroscopic $T_{\rm eff}$ can be obtained from the InfraRed Flux Method (IRFM)  \citep{blackwell&shallis1977}. There is no sign of any interstellar Na\:D lines in the spectrum, so reddening is expected to be negligible. Broad-band APASS photometry from UCAC4 \citep{zachariasetal2012} and 2MASS were used to estimate the observed bolometric flux. The IRFM was then used to determine $T_{\rm eff} = 7110 \pm 150$\,K, which is consistent with the $T_{\rm eff}$ from the spectral analysis.

As another check, we also calculated a quartic polynomial fit in ($g^{\prime}-r^{\prime}$) colour to calculate a temperature of 7170\,K, but the uncertainty on this value is large at about $\pm$250\,K \citep{uytterhoevenetal2011}.

We used the spectrum to discern the star's kinematics, finding a velocity shift of $-300$\,km\,s$^{-1}$, in agreement with \citet{balona&nemec2012}. We expect that the 16-km\,s$^{-1}$ radial velocity variations inferred from the light curve (cf. Fig.\,\ref{dopp}) would be detectable from spectroscopic observations that were suitably spaced in time.

Underabundances of Fe-peak elements by factors of 10--100 are characteristic of the $\lambda$\,Boo stars, which comprise only $\sim$2\:per\:cent of objects in the $\delta$\,Sct spectral region \citep{paunzenetal2002}. If this star is determined through a full abundance analysis to be of the $\lambda$\,Boo class, it will be the most intensely photometrically studied $\lambda$\,Boo star to date -- only eight spectroscopic binary systems with suspected $\lambda$\,Boo components are currently known \citep{paunzenetal2012}. With our low S/N spectrum it is not possible to determine whether C, N and O are normal, as in the $\lambda$\,Boo stars, but many elements seem underabundant with respect to the Sun. The observed abundances are entirely consistent with the classification of KIC\,11754974 as an SX\,Phe star by \citet{balona&nemec2012}.

\section{Pulsation characterisation}
\label{photometry}
\subsection{The data and preliminaries}

In this paper we analysed the short cadence Q6+Q7 light curves from KASOC\footnote{\url{http://kasoc.phys.au.dk}} (the Kepler Asteroseismic Science Operations Center) for the HADS star KIC\,11754974, to which we make corrections to the times to account for the light-time effects associated with being in a binary system as mentioned in \S\,\ref{sec:binarity}. We used the Pre-search Data Conditioned data of the least-squares pipeline (PDC-LS), but made checks with the Simple Aperture Photometry (SAP) data on which only basic calibration is performed in the data processing pipeline.

The PDC data are useful for their satisfactory removal of instrumental signals, including, but not limited to: heating/cooling surrounding safe-mode events, cosmic ray events, and return to scientific operating focus and temperature after monthly downlinks (Kepler Data Characteristics Handbook\footnote{\url{http://archive.stsci.edu/kepler/manuals/Data_Characteristics.pdf}}). The PDC data do not necessarily treat all jumps and outliers. For KIC\,11754974 a manual check prior to analysis showed no clear outliers in need of removal; after fitting 166 frequencies, 103 `outliers' of the 256\,514 data points were manually discarded, 100 of which were at least 1\,mmag from the fit, and the independent frequencies were thereafter improved with non-linear least-squares fitting. While in a limited number of cases the PDC data modify some stellar variability, known cases are documented in the data release notes available on the Kepler Guest Observer website\footnote{\url{http://keplergo.arc.nasa.gov/Documentation.shtml}}, and the benefits of using PDC data over SAP data outweigh the detriments in this case. Further discussion on the implications to asteroseismology of choosing which of the SAP or PDC data to use can be found in \citet{murphy2012a, murphy2012b}.

We also note that the variability of KIC\,11754974 was already discovered from the All Sky Automated Survey (ASAS) observations (\citealt{pigulskietal2009}, star \#143 in the catalogue). Due to the quality of the data, only the main mode was detected; we did not use those data in this investigation.

The data were imported into {\small PERIOD04}, for its useful graphical user interface, convenient least-squares and discrete Fourier transform functions, and tools for frequency extraction with prewhitening. A review of the functions of {\small PERIOD04} has been published by the program's creators \citep{lenz&breger2004}, and a more comprehensive view of its capabilities can be found in the user guide \citep{lenz&breger2005}.

The corrected, edited Q6.1--Q7.3 dataset contains 256\,411 points taken from (BJD) 2455372.439 to 2455552.558 (180.12\,d), having a 97\:per\:cent duty cycle.

\subsection{The light curve}

\begin{figure*}
\begin{center}
\includegraphics[width=0.95\textwidth]{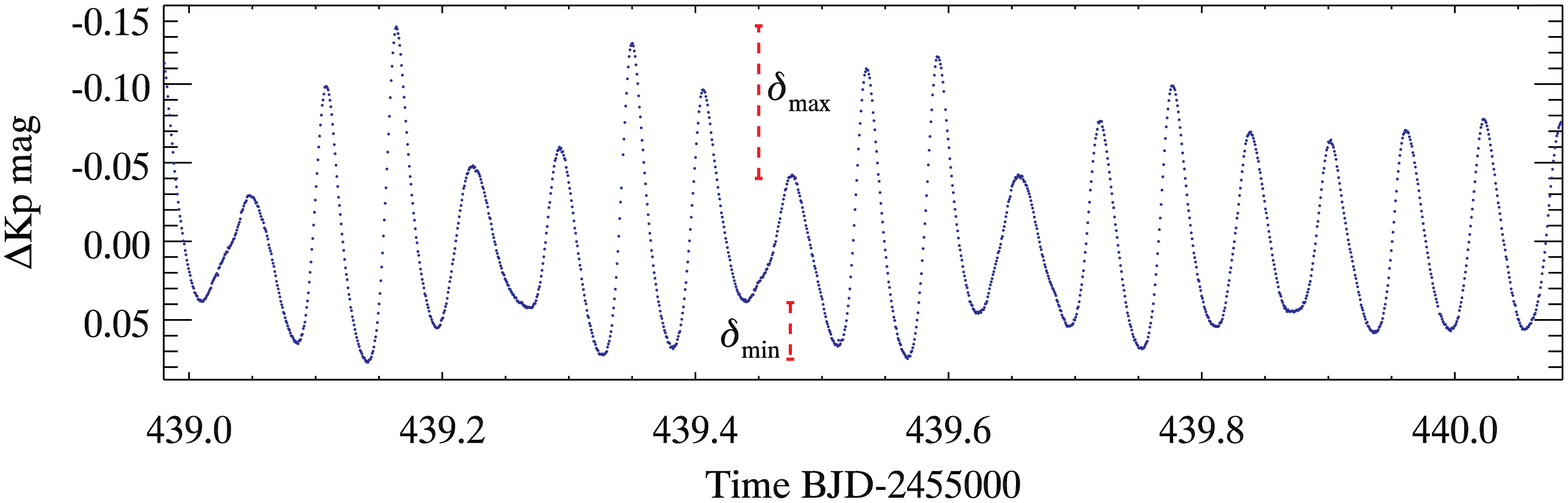}
\includegraphics[width=0.95\textwidth]{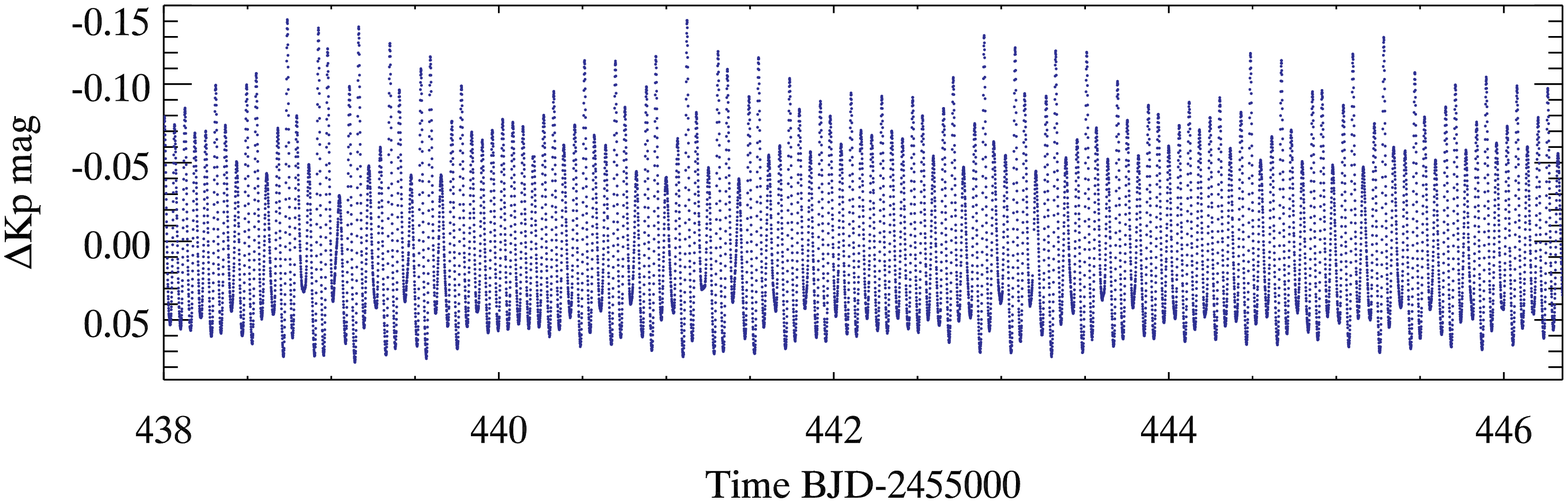}
\caption{1.1-d segment of the light curve of KIC\,11754974, showing obvious beating. Flux maximum occurs over a greater range in brightness than does flux minimum, as indicated by the fact that $\delta_{\rm max} > \delta_{\rm min}$. The lower panel indicates this is not just localised to a short time segment.}
\label{fig:lightcurves}
\end{center}
\end{figure*}

We extracted 166 frequencies that are an excellent fit to the light curve. We recalculated our fit both after outlier extraction, and after removing a cubic spline fit. The latter was done by calculating the average residuals at 2-d intervals, interpolating with a cubic spline, and subtracting the smoothed curve from the original data. Semi-regular cycles (Fig.\,\ref{fig:lightcurves}, lower panel) are evident on different timescales as a result of beating between frequencies.

The peaks in brightness occur over a greater intensity range than the troughs (Fig.\,\ref{fig:lightcurves}, upper panel). This kind of light curve shape is typical for HADS stars and can be thought of as saturation of the driving mechanism \citep{balonaetal2011b}. It has also been observed in many $\gamma$\,Dor stars, and is common in RRab (RR Lyrae) variables. In particular, it is common in white dwarfs that show many combination frequencies in their pulsation spectra. \citet{wu2001} described how in white dwarfs, which pulsate in g-modes, the convection turnover time is inferred from the phase of a combination frequency relative to its parents, and how the amplitude difference between the sums and differences in combination terms (e.g. $f_1 + f_2$ and $f_1 - f_2$) can give the thermal constant of the stellar convection zone at equilibrium. However, in an investigation into combination frequencies in $\delta$\,Sct stars, \citet{balona2012} concluded that there is no information in the relative amplitude of a combination frequency that might be useful for mode identification in these p-mode pulsators. 

\subsection{The region of independent frequencies}
\label{ssec:independent_freqs}
In KIC\,11754974 all of the high-amplitude independent frequencies are found in the range 16 to 25\,d$^{-1}$ (a full list of extracted frequencies with their corresponding amplitudes, along with identifications, is given in Table\,\ref{tbl:frequency}). Outside of this range, most frequencies detected can be explained as combination of dominant frequencies in the independently excited region. Such is also the case for the star 44\,Tau, on which the literature is extensive. In KIC\,11754974, independent peaks outside the 16 to 25\,d$^{-1}$ range have low amplitudes, typically of order 200\,$\mu$mag.

\begin{figure*}
\begin{center}
\includegraphics[width=0.98\textwidth]{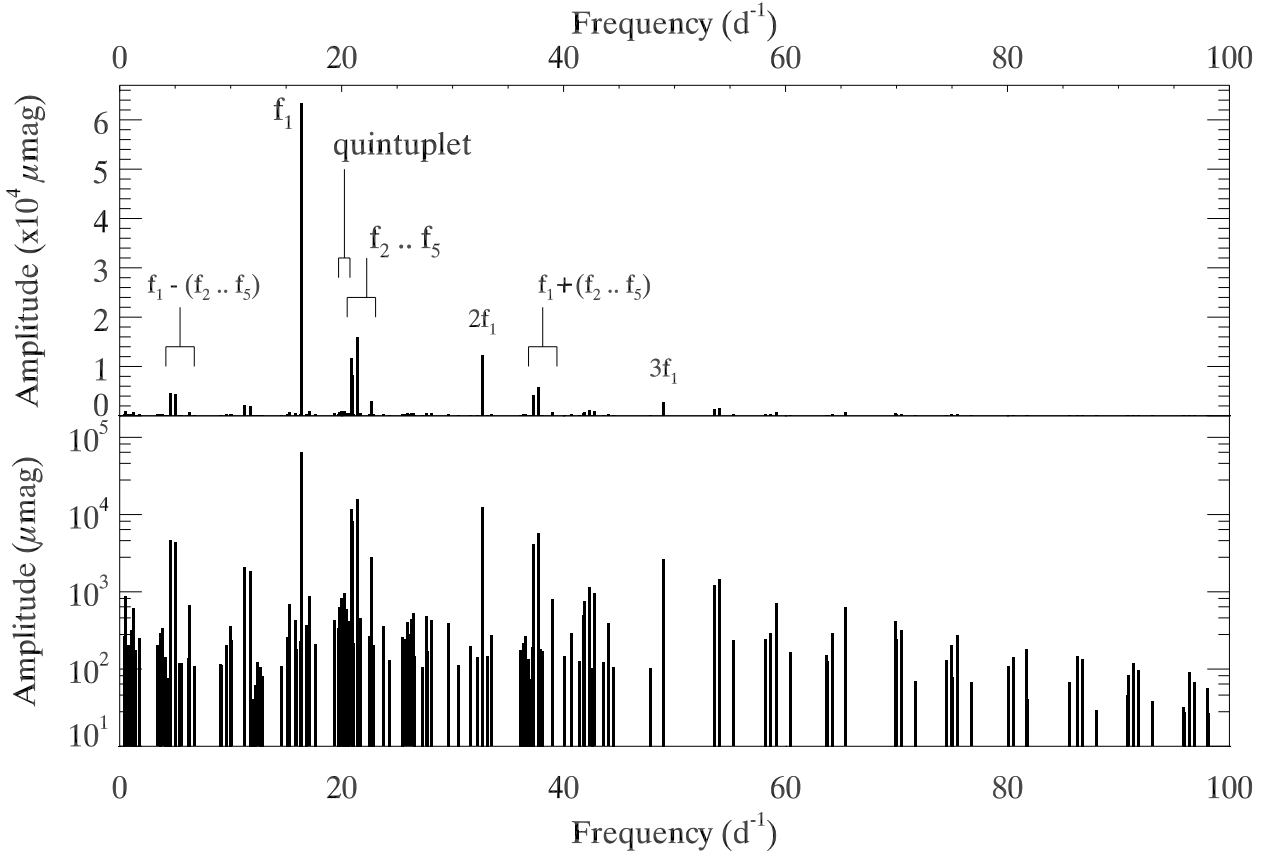}
\caption{\textit{Upper panel}: The schematic amplitude spectrum of KIC\,11754974 from 0 to 100\,d$^{-1}$. $f_1$ to $f_5$ are the five highest amplitude peaks in the frequency region $16 < f < 25$, assigned in decreasing amplitude order. This region is magnified in Fig.\,\ref{fig:prewhitened_16-25}. The $10^4\,\mu$mag peak at 32.7\,d$^{-1}$ is 2$f_1$. Also labelled are the quintuplet (\S\,\ref{ssec:quintuplet}) and some combination frequencies (\S\,\ref{ssec:combination-frequencies}). $f_4$ is very close in frequency to $f_3$; the two are unresolvable in this figure, but $f_4$ has a higher frequency and the frequencies are resolved in the data. \textit{Lower panel}: Logarithmic version of the same plot. Note that these schematic diagrams show the frequencies listed in Table\,\ref{tbl:frequency} only.}
\label{fig:initial_spectrum}
\end{center}
\end{figure*}

\begin{figure*}
\begin{center}
\includegraphics[width=0.94\textwidth]{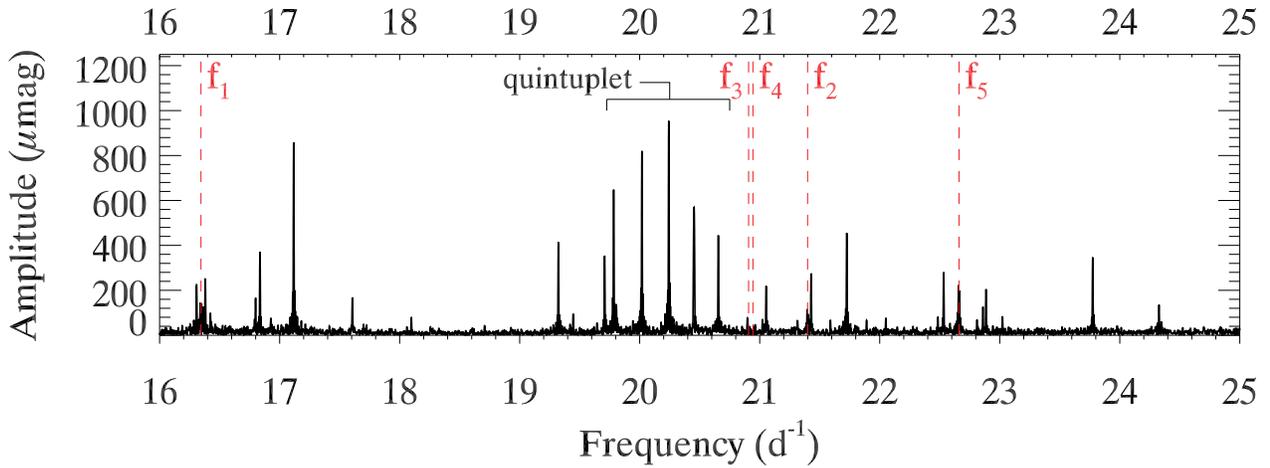}
\caption{The amplitude spectrum of the combined Q6 and Q7 dataset from 16 to 25\,d$^{-1}$ with $f_1$ to $f_5$ prewhitened. Without large amplitude modes dominating the scale, the large number of lower amplitude peaks becomes clear. The quintuplet centred at 20.243\,d$^{-1}$ is recognisable from its roughly equal spacings.}
\label{fig:prewhitened_16-25}
\end{center}
\end{figure*}

In HADS stars the period ratio of the two most dominant modes offers easy mode identification when that ratio is 0.77 (\citealt{poretti2003}, for example), as this is the theoretical period ratio of the first radial overtone mode (hereafter F1) to the fundamental radial mode (F0). The period ratio can be plotted against the base-10 logarithm of the period of F0, $\log P(F0)$, in a so-called Petersen diagram\footnote{The diagrams were first discussed by \citet{petersen1978}, but it was Art N. Cox who brought the term `Petersen diagram' into general use through his conference talks and eventually in his journal articles.}, offering diagnostic information on the star. \citet{petersen&dalsgaard1996} discuss the diagrams in detail for double-mode HADS stars of different metallicities in their figure\,3. They showed that lower values of the P(F1)/P(F0) ratio are found for metal rich stars and stars closer to the ZAMS, but spectroscopy of KIC\,11754974 indicates that this star has neither property. Hence the period ratio of its two highest-amplitude modes, $f_1$ and $f_2$, having P($f_2$)/P($f_1$) = 0.7638, is much below the anticipated value ($>0.770$) from the metallicity-calibrated Petersen diagrams. We can thus argue that $f_1$ and $f_2$ are not the F0 \textit{and} F1 modes. We also note that the high-amplitude independent frequencies around 21\,d$^{-1}$ (Figs.\,\ref{fig:initial_spectrum}\,\&\,\ref{fig:prewhitened_16-25}), particularly $f_3$ and $f_4$, could each potentially be the first radial overtone based on period ratios, though not all can be $\ell = 0$ modes, and none exactly gives the anticipated ratio.

The high amplitude of $f_1$ alone does not necessarily imply it is the fundamental radial mode, either; in FG\,Vir whose pulsation amplitudes are comparable to this star, the fundamental radial mode was not found to be the highest in amplitude of the observed frequencies \citep{guziketal2000, bregeretal2005}. 4\,CVn is another example \citep{bregeretal2008, castanheiraetal2008}. The high amplitude of $f_1$ does have other consequences: its window pattern dominates the entire analysed spectrum, even far from its frequency of 16.345\,d$^{-1}$. Peaks arising solely from the spectral window of $f_1$ have amplitudes of 0.9 and 0.5\,mmag at 50 and 100\,d$^{-1}$ respectively. To make other peaks with lower amplitudes visible, Fig.\,\ref{fig:initial_spectrum} shows a schematic amplitude spectrum with all window patterns removed. The five highest amplitude peaks in the range 16 to 25\,d$^{-1}$ are all independent modes (note that the third and fourth highest peaks in that range are indistinguishable in Fig.\,\ref{fig:initial_spectrum}). Other peaks with high amplitudes found outside this region are combinations of these frequencies.

Only once the five main independent modes are removed is it possible to see the high frequency density in the Fourier transform. Fig.\,\ref{fig:prewhitened_16-25} allows a clearer view of the lower-amplitude peaks surrounding the main pulsation frequencies. In this frequency-range independent frequencies are common, and few frequencies appear to be combinations. The spacing of 0.036\,d$^{-1}$ between $f_3$ and $f_4$ is found repeated in combination frequencies all over the spectrum, with different coefficients of $f_1$ and $f_2$ added to them.

\subsection{Quintuplet}
\label{ssec:quintuplet}

\begin{figure}
\begin{center}
\includegraphics[width=0.49\textwidth]{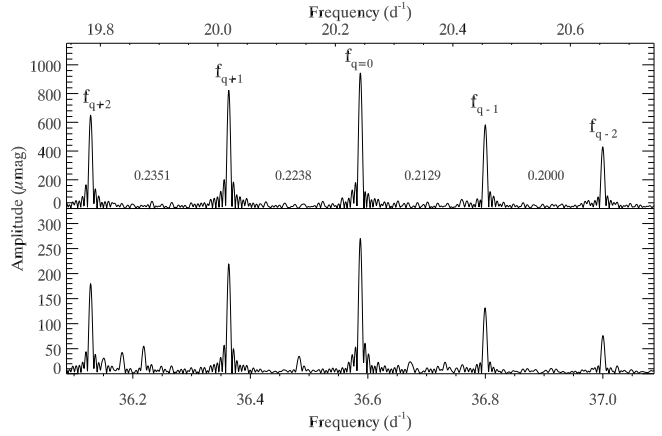}
\caption{The main quintuplet at 20.242\,d$^{-1}$ (upper panel) and the quintuplet formed in combination with $f_1$ at 36.587\,d$^{-1}$ (lower panel). The spacings in the upper panel are equal to those in the lower one because the frequencies are formed in exact combinations with $f_1$ (but are only forced to be exact in the fitting -- here we present them as they are observed). Subscripted numbers represent the $m$-values of the mode. Least-squares uncertainties on the frequencies are $<$10$^{-4}$\,d$^{-1}$ in the bottom panel and $\ll$10$^{-4}$\,d$^{-1}$ in the top panel.}
\label{fig:quintuplet}
\end{center}
\end{figure}

The Fourier transform of the light curve of the star features a quintuplet, with a central frequency at 20.242\,d$^{-1}$, and two lower-amplitude companions on each side with a mean separation of 0.218\,d$^{-1}$, shown in Fig.\,\ref{fig:quintuplet} (upper panel). We assume that the quintuplet is rotationally split. The exact morphology of the quintuplet depends on excitation and orientation, but unless one can safely assume components are excited to the same amplitude, the amplitude ratios cannot be used to constrain the star's inclination. The separations are not perfectly identical between components, nor do we expect them to be for rotationally split quintuplets -- exact frequency spacing is not expected unless there is frequency locking due to resonance \citep{buchleretal1997}. The need for high-dispersion spectroscopy or multicolour photometry is highlighted in this mode identification attempt -- better spectroscopy can more tightly constrain both $v$\,sin\,$i$ and the fundamental parameters required to improve the models (described in Section\,\ref{sec:models}), allowing more certain mode identification.

In other case studies, peaks due to rotation have been found at low frequency with highly-significant amplitudes (e.g. KIC\,9700322 \citealt{bregeretal2011, guzik&breger2011}). Hence we searched the Q6 and Q7 SC data for a peak at low frequency that might correspond to the rotation frequency. There is no evidence for a peak at the anticipated value, nor at half or double that value (cf. \citealt{balona2011}). In fact, there is no evidence for a peak due to rotation below $\sim$0.8\,d$^{-1}$, which has implications for our modelling discussion \S\,\ref{ssec:model-discussion}. To check that such a peak was not somehow removed by either our spline fit or the PDC pipeline, the SAP data were also searched for the same peak, but no peak was found.

\begin{figure}
\begin{center}
\includegraphics[width=0.50\textwidth]{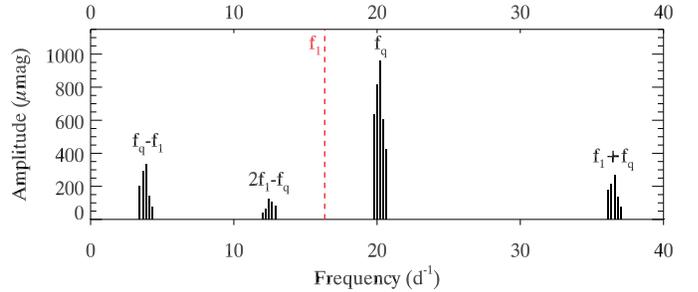}
\caption{Schematic diagram of the many quintuplets in the data. The main quintuplet at 20.242\,d$^{-1}$ forms other quintuplets throughout the Fourier spectrum in combination with $f_1$. One can see the similar amplitude ratios of the quintuplets as the shape of the quintuplet is maintained in each combination.}
\label{fig:many_quintuplets}
\end{center}
\end{figure}

Another quintuplet occurs at $\sim36.59$\,d$^{-1}$ (Fig.\,\ref{fig:quintuplet}, lower panel), but this is a manifestation of the main quintuplet in combination with $f_1$. This is confirmed by inspecting the amplitudes of this second multiplet in light of the first -- not only are they smaller, but they retain similar amplitude ratios between peaks within the multiplet. Similar quintuplets can be identified at 3.90\,d$^{-1}$, arising from $f_{\rm quintuplet} - f_1$, and at 12.45\,d$^{-1}$ from $2f_1 - f_{\rm quintuplet}$. The frequency distribution and the amplitudes of the quintuplets are shown in Fig.\,\ref{fig:many_quintuplets}.

\subsection{Combination frequencies}
\label{ssec:combination-frequencies}

The frequencies in Table\,\ref{tbl:frequency} have been extracted down to an amplitude level of 100\,$\mu$mag for the region 0--70\,d$^{-1}$, and to 20\,$\mu$mag for 70--100\,d$^{-1}$ for closer examination of patterns. The region 70--100\,d$^{-1}$ shows the amplitude level at which pulsation-mode candidates can be identified. \textit{Kepler} SC data allow high frequency analysis, and the high precision of the satellite's photometry, along with its position outside of Earth's atmosphere, afford low noise levels ($<5$\,$\mu$mag for the region 70--100\,d$^{-1}$). The combination of the two make the detection of high order frequency combinations possible.

The tolerance criterion for each calculated combination frequency has been chosen as 0.001\,d$^{-1}$; to be identified as a combination, the calculated frequency must agree with the observed frequency within this value. This tolerance is strict by some definitions -- for the dataset analysed, $1/T = 0.0056$\,d$^{-1}$ and the full width of a peak at half maximum (FWHM) is 0.0067\,d. A stricter tolerance reduces the probability of false positives for frequency combinations, and has been made necessary because of the high number of frequencies extracted.

In addition to meeting the tolerance criterion, the combination must also be physically sensible -- high coefficients and many combining independent frequencies require increasingly unlikely interactions between modes. Hence both the coefficients of the independent frequencies and the number of independent frequencies must be small. Sensible combinations typically have neither of those numbers greater than three, but no hard-and-fast rule was applied. In the region above 70\,d$^{-1}$, the coefficients have to be quite high, especially since $f_1$ is only 16.345\,d$^{-1}$. We note that some of the remaining `independent' frequencies without identifications in Table\,\ref{tbl:frequency} can be explained with physically sensible combination frequencies if the tolerance is relaxed.

{\small PERIOD04} is capable of testing for frequency combinations to a given tolerance criterion; such a method was employed by \citet{bregeretal2005} in their analysis of the star FG\,Vir, where their tolerance criterion was 0.001\,d$^{-1}$. We used the same tolerance criterion and ran simulations to determine the number of accidentally-matched frequencies one can expect. Using the frequencies $f_1$ to $f_5$, allowing coefficients between $-1$ and $6$, and allowing all five frequencies to be involved in the combinations simultaneously (i.e. there is much more freedom than in the combinations in Table\,\ref{tbl:frequency}), we find only 1--2 frequencies in the range 70 to 100\,d$^{-1}$ to be accidentally matched. Hence almost all combinations presented in Table\,\ref{tbl:frequency} are probably genuine. The two frequencies with the greatest uncertainties lie at 91.377 and 95.939\,d$^{-1}$, with errors of 0.001\,d$^{-1}$, which is the same variation as the combination frequency tolerance. These are the lowest-amplitude frequencies extracted, and are the two prime candidates for mis-matched frequencies.

\begin{figure}
\begin{center}
\includegraphics[width=0.48\textwidth]{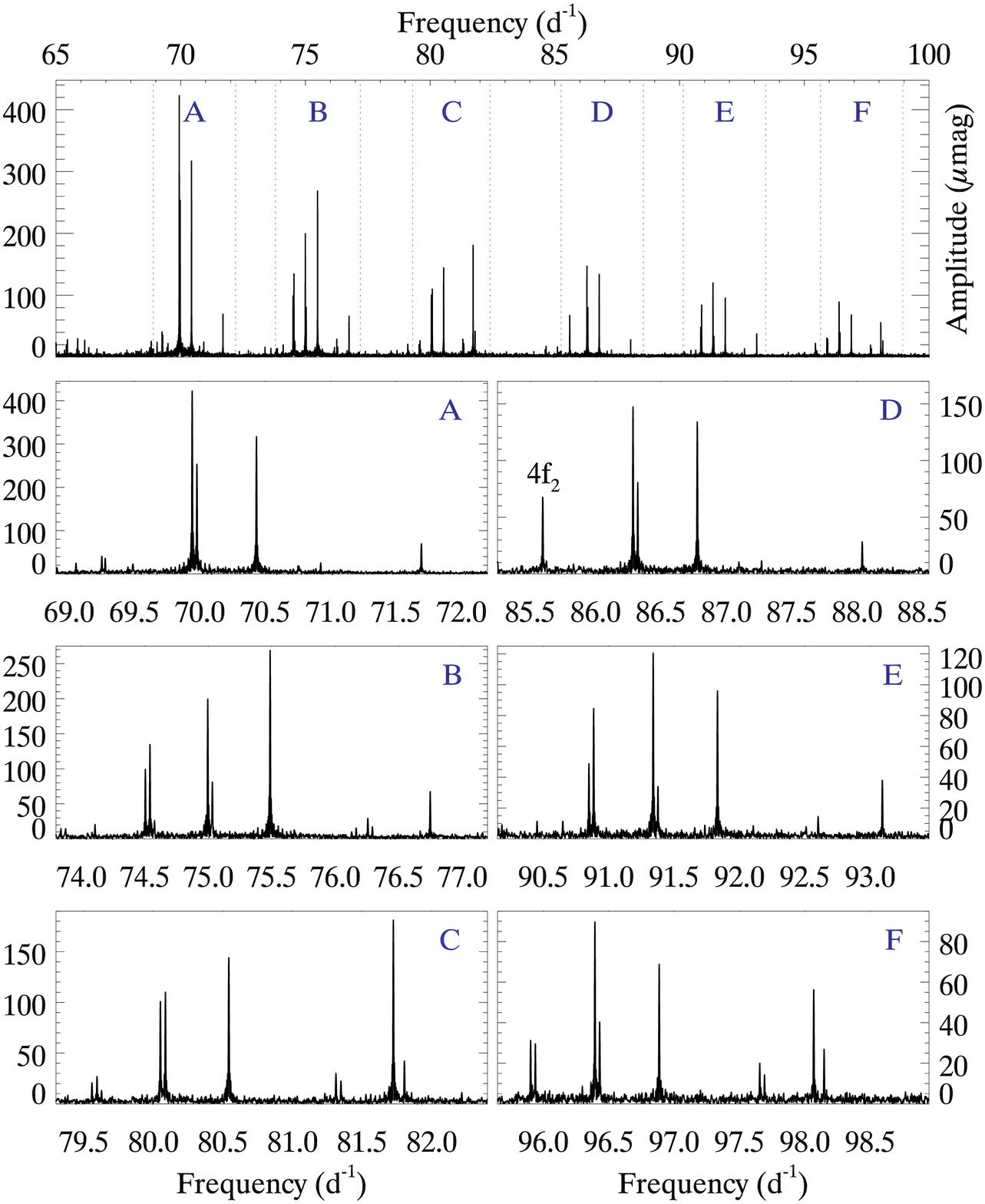}
\caption{Repetitive patterns are caused by adding $f_1$ to other combinations of frequencies, hence panels A\&D, B\&E, and C\&F are each separated in frequency by $f_1 = 16.345$\,d$^{-1}$. The frequency range displayed by each is shown in the top panel. The first peak in panel D is $4f_2$; $4f_2 - f_1$ is too low in amplitude to be distinguishable from noise, hence it is not seen in panel A. The cause of relative amplitude changes of neighbouring peaks following the addition of $f_1$ is unclear, but the absolute changes in amplitudes in adjacent \textit{panels} (note the change in axis scale -- units are $\mu$mag) is a result of having a greater number of combination terms.}
\label{fig:patterns}
\end{center}
\end{figure}

Combination frequencies are even detected at beyond 100\,d$^{-1}$ (not shown). Particular caution has been made in identifying high frequencies as combinations, especially because they often contain more than two independent frequencies and coefficients of 3, 4 or even higher. There is no doubt however, that these are indeed combinations. Several patterns of frequencies repeat themselves, separated by $f_1$ (Fig. \ref{fig:patterns}), in a way that could only happen dependently. The frequencies of these peaks are entirely calculable given the frequencies of the main independent modes, and at yet higher frequencies they recur with diminishing amplitudes until $\sim$140\,d$^{-1}$ at which point their amplitudes become indistinguishable from noise and the patterns become less discernible when peaks are missing.

\subsection{Validity of extracted frequencies}

\begin{figure}
\begin{center}
\includegraphics[width=0.5\textwidth]{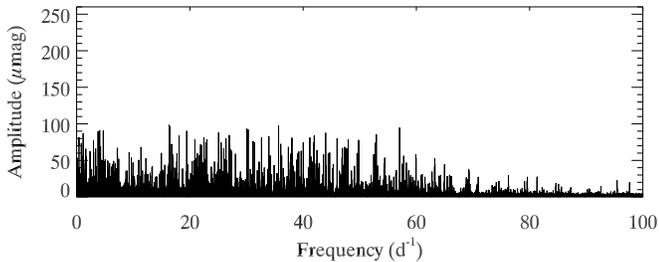}
\caption{Residuals left behind after extraction of 166 frequencies, to amplitude limits $100\,\mu$mag for $0 \leq f \leq 70$\,d$^{-1}$, and $20\,\mu$mag for $70 < f \leq 100$\,d$^{-1}$.}
\label{fig:residuals}
\end{center}
\end{figure}

One could continue to extract frequencies down to lower and lower amplitudes, but this begs the question of where to stop. 100\,$\mu$mag was chosen as a sensible limit, to extract the important frequencies for mode identification, and also for clarity: mode splittings become harder and harder to spot when the density of peaks increases, as happens when one lowers the extraction limit. On occasion, combination frequencies with amplitudes below 100\,$\mu$mag were extracted, such as when they are expected as part of a multiplet. There are additional complications to extracting frequencies to lower amplitudes. With a 180-d dataset, frequency resolution is 0.006\,d$^{-1}$. One cannot extract frequencies more closely spaced than this. Moreover, nothing is gained by continuing extraction to lower amplitudes -- combination frequencies become less reliable as the risk of false positives rises sharply with the number of extracted frequencies, mode identification becomes more difficult when mode density increases, and the low amplitude frequencies become `noise' interspersed with the meaningful high-amplitude peaks. Fig.\,\ref{fig:residuals} shows the residuals left after pre-whitening all frequencies in Table\,\ref{tbl:frequency}.

The Kepler Input Catalogue (KIC; \citealt{brownetal2011}) provides information on the amount of light contamination appearing in the data. The value of contamination for KIC\,11754974 is 0.009, hence less than 1\:per\:cent of light incident on the CCD pixels attributed to this star can be coming from background stars. As far as we can tell from looking at \kepler\ Full Frame Images, 2MASS and DSS images, KIC\,11754974 is safe from contamination. This makes negligible the possibility that a background star has a high-amplitude pulsation mode that is picked up and interpreted as a pulsation mode for KIC\,11754974. We know the star is in a binary system, but with a star significantly less massive than the Sun, contributing little light, and not expected to generate frequencies in the periodogram at the amplitudes and frequencies observed.

\begin{figure}
\begin{center}
\includegraphics[width=0.45\textwidth]{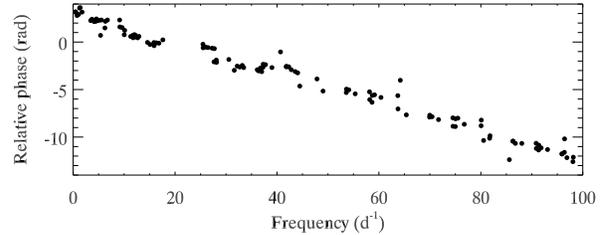}
\caption{Relative phases are negatively correlated with frequency for the combination frequencies. Those points located far from the others might be interpreted as independent frequencies instead of combinations. Error bars for phases are shown, but are generally much smaller than the plot symbols. Phases and errors are the formal least-squares values and were calculated with respect to the mid-point of the combined Q6 and Q7 SC dataset: BJD$_{\rm UTC}$ 2455462.49792.}
\label{fig:phases}
\end{center}
\end{figure}

It may be possible to check whether a frequency is a part of a combination or is an independent frequency by examining the relationship between phase and frequency for the combination terms. In a recent paper on KIC\,9700322, in which many combination frequencies were detected, \citet{bregeretal2011} found that the higher frequency modes have greater relative phases. The theory is based on that developed by \citet{buchleretal1997} who described nonlinear resonances mathematically, and \citet{balona2012} investigated further. A combination term with frequency $f = n_{\rm i}f_{\rm i} + n_{\rm j}f_{\rm j}$ should have a relative phase, $\phi_{\rm r} = \phi_{\rm c} - (n_{\rm i}\phi_{\rm i} + n_{\rm j}\phi_{\rm j})$, where $\phi_{\rm c}$ is the phase calculated by least-squares fitting. We find that $\phi_{\rm r}$ decreases with frequency once phase folding about $2\pi$ is taken into account -- we shifted groups of points by multiples of $2\pi$ in phase until they lay roughly in a straight line, then made $2\pi$ corrections for individual points lying more than $\pi$ away from the fit in an iterative way. All combination frequencies listed in Table\,\ref{tbl:frequency} are plotted in Fig.\,\ref{fig:phases}. Those `combination frequencies' that do not fit the correlation between $\phi_{\rm r}$ and frequency in the figure might actually be independent frequencies, instead. On the other hand perhaps the relationship is a purely mathematical construct. The points lying furthest from the fit are 3$f_2$ and 4$f_2$, which given the nature of the other combination frequencies in Table\,\ref{tbl:frequency} would be expected to be present and with high S/N.

The correlation has been discussed by \citet{balona2012}, who showed that for two interacting modes the phases of the harmonics of a given mode decrease linearly with increasing harmonic order (i.e. frequency). The observational evidence was not discussed in light of the precision with which frequencies, amplitudes and phases can be determined with continuous space-based observations. The effect of such small error bars is to reduce the significance of the observational evidence substantially -- the point with the median deviation lies over 27\,$\sigma$ away from the fit. The median was chosen because taking the mean is biased towards those points with minuscule relative phase errors, whose distance from the fit can be many hundred $\sigma$. The combinations giving rise to such small errors typically involve $f_1$ and $f_2$, the calculated phase errors of which are as low as 10$^{-4}$ rad. This further supports a rejection of the hypothesis of any meaningful correlation, because combinations like $f_1 + f_2$ are superiorly statistically significant in terms of S/N, and should consequently lend greater weight in this correlation. The reduced chi-squared parameter, $\chi_{\rm red}^2$, is on the order of $10^5$, clearly indicating an inappropriate fit.

To establish whether the correlation need be rejected entirely we ran a calculation whereby the frequency combinations and their phase errors were kept fixed, but instead of calculating phases from least-squares we randomised them. We found that a correlation cannot be constructed by adding $\pm n 2\pi$ to the random phases, so while it is clear that in the relative phases and frequencies of KIC\,11754974 some correlation does exist, we conclude the correlation is neither tight nor useful.

\section{Asteroseismic Models}
\label{sec:models}

KIC\,11754974 lies near the red edge of the $\delta$\,Sct instability strip. At these cooler temperatures, convection plays an important role. Indeed, without Time-Dependent Convection (TDC) treatments, an actual red edge is not accurately predicted. Thus the introduction of TDC models was an important step forward in the modelling of $\delta$\,Sct stars \citep{dupretetal2004, dupretetal2005a}. \citet{murphyetal2012} gave a successful first application of these models to an individual $\delta$\,Sct star observed by \textit{Kepler}.

Rotation plays a big role in the evolution and oscillations of $\delta$\,Sct stars \citep{goupiletal2005, rodriguezetal2006a, rodriguezetal2006b, porettietal2011}, even for KIC\,11754974, which rotates slowly compared to the distribution of $\delta$\,Sct-star rotational velocities. Furthermore, asteroseismic analyses with rotating models can assist with mode identification through the effects of rotation on the mode couplings and frequency relationships \citep{suarezetal2006, suarezetal2007}.

No combination of both TDC and rotating models exists at present, so our modelling approach consists of the two methods applied separately. This section explains the details of the models used, and the methodology followed.

\subsection{Description of the models}
\label{ssec:model-description}

Our TDC models use the same tools and selection method as that explained in detail in \citet{murphyetal2012}; the codes used are the Code Li{\'e}geois d'{\'e}volution stellaire (CL{\'E}S; \citealt{scuflaireetal2008}) for structure models and the {\small MAD} code \citep{dupret2001} supplemented with Gabriel's treatment of TDC \citep{gabriel1996, grigahceneetal2005} for non-adiabatic non-radial oscillation calculations. Convection was treated using Mixing-Length-Theory \citep[MLT]{bohm-vitense1958}, and TDC runs throughout the evolution calculations.

In order to theoretically characterise KIC\,11754974, \emph{pseudo-rotating} seismic models \citep[see][]{soufietal1998, suarezetal2002} of the star were built. These equilibrium models were computed with the evolutionary code {\sc cesam} \citep{morel1997} using frozen convection (FC) and taking the strongest effects of rotation into account. The latter is done by including the spherically-averaged contribution of the centrifugal acceleration, which is included by means of an effective gravity $g_{\mathrm{eff}}=g-{\cal A}_{c}(r)$, where $g$ is the local gravity, $r$ is the radius, and ${\cal A}_{c}(r)=\frac{2}{3}\,r\,\Omega^2(r)$ is the centrifugal acceleration of matter elements. The non-spherical components of the centrifugal acceleration are included in the adiabatic oscillation computations, but not in the equilibrium models \citep[details in][]{suarezetal2006b}.

Since we have no information about the internal rotation profile of KIC\,11754974, i.e., about how the angular momentum is distributed in its interior, we have adopted the hypothesis of differential rotation caused by local conservation of the angular momentum (in the radiative zones) during the evolution of the star. Similar rotation profiles have been found when analysing the evolution of giant stars including rotationally-induced mixing of chemical elements and transport of angular momentum \citep{maeder&meynet2004}. The physics of the equilibrium models has been chosen as adequate for intermediate-mass A-F stars, and was described in detail by \citet{suarezetal2009}.

Adiabatic oscillations were computed using the adiabatic oscillation code {\sc filou} \citep{suarez2002, suarez&goupil2008}. This code provides theoretical adiabatic oscillations of a given equilibrium model corrected up to second-order for the effects of rotation. These include near-degeneracy effects (mode coupling effects), which occur when two or more frequencies are close to each other. In addition, the perturbative description adopted takes radial variation of the angular velocity (differential rotation) into account in the oscillation equations. The oscillation spectra were calculated from frequencies around the fundamental radial mode (for each model) and the cut-off frequency. This allows the presence of low-order g and p modes (mixed modes), which are generally present in $\delta$\,Sct stars. Since non-radial oscillations are expected, modes were computed from degrees $\ell=0$ (radial) to $\ell=3$. Although the calculations are adiabatic, the effects of rotation alter the frequencies by an amount greater than the use of non-adiabatic codes.

\subsection{Methodology}
\label{ssec:methodology}

The low $T_{\rm eff}$ of the target suggested TDC models would provide a more accurate first approach to the global properties of the star. We selected the best-fitting model by perturbing the atmospheric parameters from \S\,\ref{sec:spectroscopy} (Table\,\ref{spectroscopic-parameters}) to obtain the best match with the observed oscillation frequencies, adjusting the TDC parameters (namely the convection efficiency) accordingly. The properties of the best TDC model are presented in Table\,\ref{TDC-model-properties}.

It is important to stress that the TDC models do not include rotation. Our next step was therefore to calculate equivalent (in terms of global parameters) FC models, incorporating rotation. We considered that the error committed this way (i.e. TDC first) is much lower than through direct modelling with FC and rotation.

The modelling with rotating models involves building a grid, with masses ranging from 1.4 to 2.0\,M$_{\sun}$ in steps of 0.1\,M$_{\sun}$, a convective efficiency ranging from 0.5 (prescribed by \citealt{casasetal2006} for A-F stars) to 2.0 (which gave the best fit in TDC calculations) in steps of 0.1, and an overshoot parameter $d_{\rm ov}$ in the range of [0, 0.3] in steps of 0.05. The surface rotational velocity of the models, $v_{\rm eq}$, ranged between 20 and 30\,km\,s$^{-1}$ offering enough dispersion to analyse different configurations of mass, metallicity and age. In principle this gives splittings close to those observed for the quintuplet, but this was not constrained to be the case. In addition, it indirectly guarantees a dispersion in the inclination angle probed within the model grid. Metallicity was limited to [Fe/H]$ = -0.5 \pm 0.1$\,dex, though metallicities closer to the solar values were investigated and offered a poorer match with the observed frequencies. This implies the entire star is metal poor, and that we are not just observing a surface abundance anomaly, thus arguing against a $\lambda$\,Boo classification (cf. debates in the literature about the interior metallicity of $\lambda$\,Boo stars, e.g. \citealt{moyaetal2010}).

Model selection involves searching through the grid, minimising the mean-square error function $\chi^2$. For each mode,
\eqn{\chi^2= \frac{1}{N}\sum_{i=1}^N \Big(f_{{\rm obs},i}-\nu_{{\rm th},i}\Big)^2 \label{eq:defchi2}}
is calculated, where $f_{{\rm obs},i}$ and $\nu_{{\rm th,}i}$ represent the observed and theoretical frequencies respectively. The total number of observed frequencies is represented by $N$. Calculations are made considering the mass $M$, the rotational velocity $v_{\rm eq}$, metallicity, $\amlt$, $\dov$, and age as free parameters; pulsation amplitude information is not directly used. Importantly, all the frequencies are fitted simultaneously; assumptions like the ``strongest modes'' being radial or quintuplet membership were not used as constraints for the models.

\begin{figure}
\begin{center}
\includegraphics[width=0.49\textwidth]{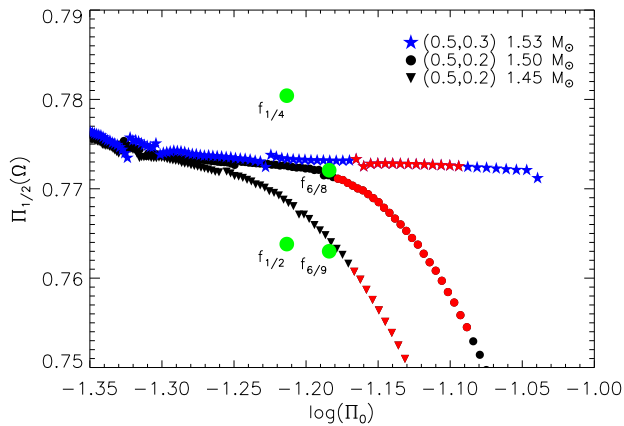}
\caption{Rotational Petersen Diagram covering models with 1.45--1.60\,M$_{\sun}$. Coloured in red are the models that satisfy the $T_{\rm eff}$ and $\log g$ uncertainties ($7000\pm200$\,K; $3.6\pm0.5$); we used a conservative $\log g$ uncertainty, because atmosphere models do not include rotation to calculate an effective gravity. Also plotted (green dots) are frequency ratios (e.g. $f_{1/4}$ indicates $f_1/f_4$) that fall within $0.772 \pm 0.009$ (see text). Numbers in brackets in the legend denote the value of $\alpha_{\rm MLT}$ and the overshooting parameter, respectively.}
\label{fig:rpd}
\end{center}
\end{figure}

We did look separately for diagnostic frequency ratios in the data. We took models of three different masses and searched for ratios of $0.772 \pm 0.009$ (cf. discussion in \S\,\ref{ssec:independent_freqs}) among the independent modes. Fig.\,\ref{fig:rpd} displays the four possible frequency ratios observed, none of which satisfies the $T_{\rm eff}$ and (rather loose) $\log g$ uncertainties, or indicates a particularly promising identification for the first radial overtone. We stress this only reduces the likelihood that F0 is present \textit{with} F1; either could still be present without the other.

\subsection{Discussion of the models}
\label{ssec:model-discussion}

In general, we found that the models with the lowest $\chi^2$ were below 1.60\,M$_{\sun}$. Moreover, those that matched either F0 or F1 to the observed frequencies had masses in the range $1.50-1.53$\,M$_{\sun}$, which are very similar to the TDC-predicted masses. The best-fitting rotating model identifies $f_1$ as the fundamental radial mode, and the quintuplet as components of a mixed ($n$, $\ell$) = ($-1$, $3$) mode. The parameters of this model are displayed in Table\,\ref{rotating-model-properties}.

That the quintuplet is identified as a mixed mode has strong implications. It directly affects the Ledoux constant, $C_L$, depending on how much g-mode character such a mode has. For lower radial orders, $C_L$ can exceed 0.1 for $\ell=2$ modes. Therefore, we had to search a range of low frequencies to account for this factor in \S\,\ref{ssec:quintuplet}, but still no peak due to rotation was found up to twice the expected value (i.e between 0.218 and 0.436\,d$^{-1}$).

The asymmetry of the quintuplet gives a proportionality constant, $D_L$, for the second order effect (centrifugal force) of rotational splitting, which we can use as a consistency check. For moderately-rotating $\delta$\,Sct stars,
oscillations can be expressed in terms of the perturbation theory as $$\omega = \omega_0 + (C_L - 1)m\Omega + D_L\frac{m^2 \Omega^2}{\omega_0},$$ which describes the effects of rotation up to second order \citep[adapted from][]{dziembowski&goode1992}, for a rotation frequency, $\Omega$, and observed and intrinsic (unperturbed) pulsation frequencies, $\omega$ and $\omega_0$, respectively. Under the assumption that the quintuplet is actually an $\ell=2$ mode, one gets $D_L$=2.45 in the case of $C_L$=0, or $D_L$=1.8 for $C_L$=0.12. These are about the expected values for an ($n$, $\ell$) = ($2$, $2$) mode. (See \citealt{dziembowski&goode1992} for the example values). The Q-value of an ($n$, $\ell$) = (2,2) mode at the observed quintuplet frequencies is consistent with the assumption that the lowest independent frequency is near the radial fundamental mode.

Thus we have the following scenario: the best rotating model indicates that $f_1$ is the fundamental radial mode, and inferences on the Q-value of the quintuplet if it is an ($n$, $\ell$) = (2,2) mode, would agree with that identification for $f_1$. However, the model predicts that the quintuplet is part of an $\ell=3$ mode instead. This is only an apparent disagreement. First, that an ($n$, $\ell$) = (2, 2) mode can generate the observed frequency quintuplet does not make its identification as part of an $\ell=3$ septuplet impossible, as the expected rotational splitting asymmetry of such a mode is expected to be quite similar. Second, differential rotation in the radial direction is another possible explanation. It has been proven that the oscillations of low-order p and g modes are highly sensitive to variations of the rotation profile near the core \citep{suarezetal2006}. Further attempts at direct mode identification, through multi-colour photometry, and further constraints for the models from high-S/N, high-resolution spectroscopy are required for further investigation.

\section{Conclusions}

KIC\,11754974 is a cool, high-amplitude $\delta$\,Sct star with a large number of combination frequencies that show a predictable pattern at high frequency. We discovered from the pulsations that the star is in a 343-d binary system; our seismic models show the primary (pulsating) member is 1.53\,M$_{\sun}$, making the secondary 0.63\,M$_{\sun}$ for a model-determined $i=47^{\circ}$. No direct contribution from the secondary is seen in the spectrum or the light curve.

Compared to most A-type stars, KIC\,11754974 is a slow rotator, which is typical of HADS stars, and allowed us to construct perturbative rotating models. The models indicate the star is metal-poor, in that solar-metallicity models were a poor fit to the observed frequencies. This agrees with literature classifications that this is a Population\,II (SX\,Phe) star, and argues against a surface abundance anomaly model like the Am stars or $\lambda$\,Boo stars.

The location of this star on the rotating Petersen diagram argues against more than one of the high-amplitude $f_1$..$f_5$ modes being radial. We cannot confirm whether the quintuplet found in the data is rotationally split -- direct mode identification is required for this.

Combination frequencies are present all across the spectrum, arising from five independent modes that are all confined to a small frequency region. Outside the frequency region $16<f<25$\,d$^{-1}$ independent modes with amplitudes above 500\,$\mu$mag are non-existent, and most frequencies are combinations of some form.

The star has a lot of asteroseismic potential, being near the red-edge of the $\delta$\,Sct instability strip where the pulsation-convection interaction is under study, being a Population\,II star with precise and sensitive photometry, and being a high amplitude pulsator for which mode interaction is also particularly important. Here we have scratched the surface with our models, but there is great potential to learn much more. We intend to study the star further, with multicolour photometry and higher-resolution spectroscopy to identify the pulsation modes and constrain the atmospheric parameters. Despite being faint, the star is a promising target for learning about mode selection in these pulsators.

\section*{Acknowledgements}
This paper includes data collected by the \textit{Kepler} mission. Funding for the \textit{Kepler} mission is provided by the NASA Science Mission directorate. The paper also includes observations made with the Nordic Optical Telescope, operated jointly by Denmark, Finland, Iceland, Norway, and Sweden, on the island of La Palma at the Spanish Observatorio del Roque de los Muchachos of the Instituto de Astrof\'{\i}sica de Canarias (IAC). SJM would like to acknowledge the financial support of the STFC, and from the projects FCOMP-01-0124-FEDER-009292 \& PTDC/CTE-AST/098754/2008 under grant CAUP-09/2012-BI. AP acknowledges the support from the NCN grant No. 2011/03/B/ST9/02667. Some calculations have been carried out in Wroc{\l}aw Centre for Networking and Supercomputing (\url{http://www.wcss.wroc.pl}), grant No. 219. JCS acknowledges support by 
Spanish National Research Plan through grants ESP2010-20982-C02-01 and  AYA2010-12030-E. KU acknowledges financial support by the Spanish National Plan of R\&D for 2010, project AYA2010-17803. RS was supported by the J\'anos Bolyai Research Scholarship, the `Lend\"ulet-2009 Young Researchers' Program of the Hungarian Academy of Sciences, the HUMAN MB08C 81013 grant of the MAG Zrt, the Hungarian OTKA grant K83790 and the European Community's Seventh Framework Programme (FP7/2007-2013) under grant agreement no. 269194 (IRSES/ASK). AOT acknowledges support from Sonderforschungsbereich SFB 881 ``The Milky Way System'' (subproject A5) of the German Research Foundation (DFG). GH is thankful for support by the NCN grant 2011/01/B/ST9/05448.

\bibliography{arxiv_11754974}

\clearpage

\begin{table*}
\begin{center}
\caption{Multifrequency extraction and identification for KIC\,11754972. The absence of combinations between 16 and 25\,d$^{-1}$ is evident; the quintuplet and $f_1$ to $f_5$ all lie in this region. Frequencies are only given an individual identification if they are of particular significance or form a combination. Errors on frequencies are generally smaller than 2x10$^{-4}$\,d$^{-1}$, and errors on amplitudes are all under 4\,$\mu$mag. For $f_1$ to $f_5$, from which combinations are calculated, the frequency uncertainty is 10$^{-6}$\,d$^{-1}$. * denotes quintuplet membership. $\dagger$ this frequency is not fully resolved from $f_2$ in the SC data.}
\begin{tabular}{r r c r r r c r r r c}
\hline
frequency & amp. & ID & \hspace{5mm} & frequency & amp. & ID & \hspace{5mm} & frequency & amp. & ID \\
d$^{-1}$ & $\mu mag$ & & & d$^{-1}$ & $\mu mag$ & & & d$^{-1}$ & $\mu mag$ & \\
\hline
$	16.34474	$ & $	63329	$ & $	f_1	$ &		&		&		&		&		&		&		&			\\
$	21.39898	$ & $	15940	$ & $	f_2	$ &		&		&		&		&		&		&		&			\\
$	20.90740	$ & $	11653	$ & $	f_3	$ &		&		&		&		&		&		&		&			\\
$	20.94355	$ & $	8229	$ & $	f_4	$ &		&		&		&		&		&		&		&			\\
\vspace{2mm}$	22.66020	$ & $	2839	$ & $	f_5	$ &		&		&		&		&		&		&		&			\\
$	0.4554	$ & $	270	$ & $	=f_2-f_4	$ &		& $	20.4552	$ & $	603	$ & $	=f_{11}*	$ &		  & $	42.7980	$ & $	946	$ & $	=2f_2	$	\\
$	0.4916	$ & $	865	$ & $	=f_2-f_3	$ &		& $	20.6552	$ & $	422	$ & $	=f_{12}*	$ &		& $	43.5676	$ & $	122	$ & $	=f_3+f_5	$	\\
$	0.7723	$ & $	202	$ & $	=f_7-f_1	$ &		& $	21.0537	$ & $	216	$ & $		$ &		        & $	44.0592	$ & $	389	$ & $	=f_2+f_5	$	\\
$	1.0705	$ & $	315	$ & $	=f_1-f_6	$ &		& $	21.3973	$ & $	257	$ & $	 \dagger $ &		        & $	44.4716	$ & $	107	$ & $	=4f_1-f_3	$	\\
$	1.2612	$ & $	619	$ & $	=f_5-f_2	$ &		& $	21.4282	$ & $	274	$ & $		$ &		        & $	47.8522	$ & $	103	$ & $	=3f_2-f_1	$	\\
$	1.3804	$ & $	178	$ & $	=f_2-f_9	$ &		& $	21.7251	$ & $	458	$ & $	=f_{14}	$ &		  & $	49.0342	$ & $	2677	$ & $	=3f_1	$	\\
$	1.7528	$ & $	250	$ & $	=f_5-f_3	$ &		& $	22.5325	$ & $	270	$ & $		$ &		        & $	53.5969	$ & $	1230	$ & $	=2f_1+f_3	$	\\
$	3.4387	$ & $	202	$ & $	=f_8-f_1	$ &		& $	22.8592	$ & $	107	$ & $		$ &		        & $	53.6330	$ & $	828	$ & $	=2f_1+f_4	$	\\
$	3.6738	$ & $	291	$ & $	=f_9-f_1	$ &		& $	22.8860	$ & $	205	$ & $		$ &		        & $	54.0885	$ & $	1466	$ & $	=2f_1+f_2	$	\\
$	3.8976	$ & $	334	$ & $	=f_{10}-f_1$ &	& $	23.7744	$ & $	354	$ & $=f_{15}$ &		        & $	55.3497	$ & $	235	$ & $	=2f_1+f_5	$	\\
$	4.1040 	$ & $141 	$ & $=f_{13}-f_1$ &  & $	24.3270	$ & $	130	$ & $		$ &	      	  & $	58.1595	$ & $	246	$ & $	=f_1+2f_3	$	\\
$	4.1105	$ & $	161	$ & $	=f_{11}-f_1	$ &	& $	25.4701	$ & $	257	$ & $	=2f_3-f_1	$ &	  & $	58.1957	$ & $	174	$ & $	=f_1+f_3+f_4	$	\\
$	4.3108	$ & $	76	$ & $	=f_{12}-f_1	$ &	& $	25.5062	$ & $	229	$ & $	=f_3+f_4-f_1$&  & $	58.6511	$ & $	292	$ & $	=f_1+f_2+f_3	$	\\
$	4.5627	$ & $	4625	$ & $	=f_3-f_1	$ &	& $	25.6208	$ & $	242	$ & $		$ &		        & $	58.6873	$ & $	191	$ & $	=f_1+f_2+f_4	$	\\
$	4.5988	$ & $	2234	$ & $	=f_4-f_1	$ &	& $	25.9616	$ & $	403	$ & $	=f_2+f_3-f_1$&	& $	59.1427	$ & $	705	$ & $	=2f_2+f_1	$	\\
$	5.0542	$ & $	4435	$ & $	=f_2-f_1	$ &	& $	25.9961	$ & $	280	$ & $		$ &		        & $	60.4039	$ & $	165	$ & $	=f_1+f_2+f_5	$	\\
$	5.3804	$ & $	119	$ & $	=f_{14}-f_1	$ &	& $	26.2959	$ & $	441	$ & $	=f_{16}$ &	        	& $	63.7054	$ & $	152	$ & $	=2f_2+f_3	$	\\
$	5.5458	$ & $	118	$ & $	=2f_2-f_1-f_3$&	& $	26.4532	$ & $	524	$ & $	=2f_2-f_1	$&	  & $	63.7415	$ & $	126	$ & $	=2f_2+f_4	$	\\
$	6.2363	$ & $	139	$ & $	=3f_1-2f_2	$ &	& $	26.5422	$ & $	148	$ & $		$ &		        & $	64.1969	$ & $	292	$ & $	=3f_2	$	\\
$	6.3155	$ & $	678	$ & $	=f_5-f_1	$ &		& $	27.2590	$ & $	105	$ & $	=f_4+f_5-f_1$&	& $	65.3790	$ & $	628	$ & $	=4f_1	$	\\
$	6.7278	$ & $	108	$ & $	=3f_1-f_2-f_3$&	& $	27.6352	$ & $	481	$ & $	=3f_1-f_2$&	    & $	69.9416	$ & $	418	$ & $	=3f_1+f_3	$	\\
$	9.1253	$ & $	116	$ & $	=2f_3-2f_1	$ &	& $	27.7144	$ & $	172	$ & $	=f_2+f_5-f_1$ &	& $	69.9778	$ & $	241	$ & $	=3f_1+f_4	$	\\
$	9.1615	$ & $	112	$ & $	=f_3+f_4-2f_1$&	& $	28.0907	$ & $	238	$ & $	=3f_1-f_4	$ &		& $	70.4332	$ & $	318	$ & $	=3f_1+f_2	$	\\
$	9.6169	$ & $	204	$ & $	=f_2+f_3-2f_1$&	& $	28.1268	$ & $	426	$ & $	=3f_1-f_3	$ &		& $	71.6944	$ & $	70	$ & $	=3f_1+f_5	$	\\
$	10.0293	$ & $	361	$ & $	=2f_1-f_5	$ &		& $	29.6549	$ & $	393	$ & $		$ &	          & $	74.5043	$ & $	94	$ & $	=2f_1+2f_3	$	\\
$	10.1085	$ & $	233	$ & $	=2f_2-2f_1	$ &	& $	29.6621	$ & $	158	$ & $		$ &	          & $	74.5404	$ & $	132	$ & $	=2f_1+f_3+f_4	$	\\
$	11.2905	$ & $	2066	$ & $	=2f_1-f_2	$ &	& $	30.5486	$ & $	112	$ & $	=2f_6	$ &		    & $	74.9959	$ & $	201	$ & $	=2f_1+f_2+f_3	$	\\
$	11.7459	$ & $	1068	$ & $	=2f_1-f_4	$ &	& $	31.6190	$ & $	196	$ & $	=f_1+f_6	$ &		& $	75.0320	$ & $	79	$ & $	=2f_1+f_2+f_4	$	\\
$	11.7821	$ & $	1844	$ & $	=2f_1-f_3	$ &	& $	32.1979	$ & $	141	$ & $	=2f_1+f_3-f_2$&	& $	75.4874	$ & $	271	$ & $	=2f_1+2f_2	$	\\
$	12.0343	$ & $	41	$ & $	=2f_1-f_{12}$&	& $	32.6895	$ & $	12311	$ & $	=2f_1	$ &		  & $	76.7487	$ & $	67	$ & $	=2f_1+f_2+f_5	$	\\
$	12.2342	$ & $	62	$ & $	=2f_1-f_{11}$&	& $	32.7252	$ & $	112	$ & $	=2f_1+f_4-f_3$&	& $	80.0501	$ & $	99	$ & $	=2f_2+f_1+f_3	$	\\
$	12.4471	$ & $	121	$ & $	=2f_1-f_{10}$&	& $	33.1811	$ & $	148	$ & $	=2f_1+f_2-f_3$&	& $	80.0863	$ & $	110	$ & $	=f_1+2f_2+f_4	$	\\
$	12.6709	$ & $	107	$ & $	=2f_1-f_9	$ &		& $	33.4618	$ & $	272	$ & $	=f_1+f_7	$ &		& $	80.5417	$ & $	142	$ & $	=3f_2+f_1	$	\\
$	12.9060	$ & $	81	$ & $	=2f_1-f_8	$ &		& $	36.1282	$ & $	175	$ & $	=f_1+f_8	$ &		& $	81.7237	$ & $	182	$ & $	=5f_1	$	\\
$	14.5919	$ & $	108	$ & $	=f_1+f_3-f_5$&	& $	36.3633	$ & $	216	$ & $	=f_1+f_9	$ &		& $	81.8029	$ & $	41	$ & $	=f_1+2f_2+f_5	$	\\
$	15.0835	$ & $	262	$ & $	=f_1+f_2-f_5$&	& $	36.5871	$ & $	270	$ & $	=f_1+f_{10}	$&	& $	85.5959	$ & $	68	$ & $	=4f_2	$	\\
$	15.2743	$ & $	692	$ & $	=f_6	$ &		    & $	36.7935	$ & $	112	$ & $	=f_1+f_{13}	$&	& $	86.2864	$ & $	146	$ & $	=4f_1+f_3	$	\\
$	15.8532	$ & $	425	$ & $	=f_1+f_3-f_2$&	& $	36.8000	$ & $	136	$ & $	=f_1+f_{11}	$&	& $	86.3225	$ & $	77	$ & $	=4f_1+f_4	$	\\
$	15.8893	$ & $	180	$ & $	=f_1+f_4-f_2$&	& $	37.0000	$ & $	74	$ & $	=f_1+f_{12}	$&	& $	86.7780	$ & $	134	$ & $	=4f_1+f_2	$	\\
$	16.3086	$ & $	230	$ & $	=f_1+f_3-f_4$&	& $	37.2410	$ & $	192	$ & $		$ &		        & $	88.0392	$ & $	29	$ & $	=4f_1+f_5	$	\\
$	16.3618	$ & $	155	$ & $		$ &		        & $	37.2521	$ & $	4143	$ & $	=f_1+f_3	$&	& $	90.8490	$ & $	46	$ & $	=3f_1+2f_3	$	\\
$	16.3807	$ & $	285	$ & $	=f_1+f_4-f_3$&	& $	37.2883	$ & $	3158	$ & $	=f_1+f_4	$&	& $	90.8852	$ & $	83	$ & $	=3f_1+f_3+f_4	$	\\
$	16.7999	$ & $	141	$ & $	=f_1+f_2-f_4$&	& $	37.7437	$ & $	5730	$ & $	=f_1+f_2	$&	& $	91.3406	$ & $	120	$ & $	=3f_1+f_2+f_3	$	\\
$	16.8363	$ & $	369	$ & $	=f_1+f_2-f_3$&	& $	37.9345	$ & $	183	$ & $	=f_5+f_6	$ &		        & $	91.3768	$ & $	32	$ & $	=3f_1+f_2+f_4	$	\\
$	17.1170	$ & $	871	$ & $	=f_7	$ &		    & $	38.0698	$ & $	170	$ & $=f_1+f_{14}$ &		& $	91.8322	$ & $	97	$ & $	=3f_1+2f_2	$	\\
$	17.6060	$ & $	212	$ & $	=f_1+f_5-f_2$&  & $	39.0049	$ & $	795	$ & $	=f_1+f_5	$ &		& $	93.0934	$ & $	38	$ & $	=3f_1+f_2+f_5	$	\\
$	19.3224	$ & $	424	$ & $		$ &	        	& $	40.1192	$ & $	146	$ & $	 =f_1+f_15	$ &		        & $	95.9033	$ & $	32	$ & $	=2f_1+f_2+2f_3	$	\\
$	19.7078	$ & $	341	$ & $		$ &	        	& $	40.6909	$ & $	295	$ & $	=f_3+f_8	$ &		& $	95.9394	$ & $	28	$ & $	=2f_1+f_2+f_3+f_4	$	\\
$	19.7835	$ & $	636	$ & $	=f_8*	$ &	    	& $	41.4176	$ & $	128	$ & $	=f_2+f_9$ &		        & $	96.3949	$ & $	91	$ & $	=2f_2+2f_1+f_3	$	\\
$	19.8051	$ & $	152	$ & $		$ &		        & $	41.8148	$ & $	494	$ & $	=2f_3	$ &		    & $	96.4310	$ & $	42	$ & $	=2f_1+2f_2+f_4	$	\\
$	20.0186	$ & $	814	$ & $	=f_9*	$ &	    	& $	41.8509	$ & $	760	$ & $	=f_4+f_3	$ &		& $	96.8864	$ & $	68	$ & $	=3f_2+2f_1	$	\\
$	20.2424	$ & $	959	$ & $	=f_{10}*	$ &		& $	42.3064	$ & $	1139	$ & $	=f_2+f_3	$ &	& $	98.0685	$ & $	57	$ & $	=6f_1	$	\\
$	20.2487	$ & $	126	$ & $		$ &		        & $	42.3427	$ & $	233	$ & $=f_2+f_4		$ &		& $	98.1476	$ & $	27	$ & $	=2f_1+2f_2+f_5	$	\\
$	20.4487	$ & $	508	$ & $	=f_{13}	$ &	    & $	42.6408	$ & $	103	$ & $	 =f_1+f_{16}$ &		        & $		$ & $		$ & $		$	\\
\hline
\hline
\end{tabular}
\label{tbl:frequency}
\end{center}
\end{table*}

\begin{table*}
\setlength{\extrarowheight}{3pt}
\centering
\caption{Properties of the best TDC model, obtained by perturbing the spectroscopic parameters and convection efficiency.}
\begin{tabular}{c c c c c c c c c c }
\hline
Mass           & Radius &  $T_{\mathrm{eff}}$&  Age & $\log g$  &   $\log ({\rm L}/\rm{L}_{\sun})$ &   X     &   Z        &   $\alpha_{\mathrm{MLT}}$      &   $d_{\mathrm{ov}}$ \\ 
M$_{\sun}$  &  R$_{\sun}$ & K           &    Myr                 &  (cgs)       &        &           &         &\vspace{1mm}\\ 
\hline
1.56 & 2.425 & 7110   & 1829   & 3.862   & 1.130  &  0.7465   & 0.0071   &2.0     &   0.2\\ 
\hline
\hline
\end{tabular}
\label{TDC-model-properties}
\end{table*}

\begin{table*}
\setlength{\extrarowheight}{3pt}
\centering
\caption{Characteristics of the best representative rotating model. The first six columns have their usual meanings, then headings are: mean density, equatorial rotation velocity, the ratio of the equatorial rotation velocity to the keplerian orbital velocity at the stellar surface, inclination of the stellar rotation axis to the line of sight, fundamental radial mode frequency f(F0), first overtone radial mode frequency f(F1), the base-10 logarithm of the period of F0, and the period ratio of F0 and F1. The model has metallicity [Fe/H]=$-0.52$, convective efficiency, $\amlt=0.5$, and overshooting $\dov=0.3$. See \S\,\ref{ssec:methodology} for model selection criteria.}
\begin{tabular}{r c c c c c c c c c c c}
\hline
Mass & Radius & $T_{\rm eff}$ & Age & $\log g$ & ${\bar \rho}$ & $v_{\rm eq}$ & $i$ & f(F0) & f(F1) & log P(F0) & P(F0)/P(F1)\\
M$_{\sun}$ & R$_{\sun}$ & K & Myr & (cgs) & & km\,s$^{-1}$ & deg & d$^{-1}$ &  d$^{-1}$ & \vspace{1mm}\\
\hline
\vspace{1mm}
$1.53	$&$	1.764	$&$	7256	$&$	1465.93	$&$	4.129	$&$	0.392	$&$	34.18	$&$	47	_{-	15	} ^{+	\phantom{1}7	}$&$	16.327	$&$	21.1	$&$	-1.213	$&$	0.773	$\\
\hline
\hline
\end{tabular}
\label{rotating-model-properties}
\end{table*}

\end{document}